\documentclass[fleqn,usenatbib,usedcolumn]{mnras}	

\usepackage[T1]{fontenc}
\usepackage{ae,aecompl}

\usepackage{graphicx}	
\usepackage{amsmath}	
\usepackage{amssymb}	
\usepackage{txfonts}
\usepackage{threeparttable}
\usepackage{natbib}
\newcommand{\apostrophe}{'}

\title[Galaxy overdensity around quasar pairs]{Overdensity of galaxies in the environment of quasar pairs}

\author[A. Sandrinelli, R. Falomo, A. Treves ]{
A. Sandrinelli$^{1,2}
$\thanks{E-mail:asandrinelli@yahoo.it }, 
R. Falomo$^{3}$,
 A. Treves$^{1,2}$, 
 R. Scarpa$^{4}$, 
 M. Uslenghi$^{5}$
 \\
$^{1}$Universit\`a degli Studi dell\apostrophe Insubria, Via Valleggio 11, I-22100 Como, Italy\\
$^{2}$INAF -  Istituto Nazionale di Astrofisica, Osservatorio Astronomico di Brera, Via Emilio Bianchi 46, I-23807 Merate, Italy\\
$^{3}$INAF -  Istituto Nazionale di Astrofisica, Osservatorio Astronomico di Padova, Vicolo dell\apostrophe Osservatorio 5, I-35122 Padova, Italy\\
$^{4}$IAC - Instituto de Astrofisica de Canarias, C/O Via Lactea, s/n E38205 - La Laguna, Tenerife, Espa\~{n}a\\
$^{5}$INAF-IASF - Istituto Nazionale di Astrofisica, Istituto di Astrofisica Spaziale e Fisica Cosmica, via E. Bassini 15, I-20133 Milano, Italy
}

\date{Accepted XXX. Received YYY; in original form ZZZ}

\pubyear{2015}

\begin{document}
\label{firstpage}
\pagerange{\pageref{firstpage}--\pageref{lastpage}}
\maketitle

\begin{abstract}
We report on a study of the galaxy environments of low redshift physical  quasars pairs.  
We selected 20  pairs having projected separation $<$ 0.5 Mpc and difference of systemic 
velocity $<$ 800 km s$^{-1}$. Using Sloan Digital Sky Survey images we evaluated the galaxy overdensity around 
these quasars in pairs and then compare it with that of a sample of isolated quasars with same 
redshift and luminosity.
It is found that on average there is a systematic larger overdensity of galaxies around 
quasars in pairs with respect to that of isolated quasars.  This may represent a significant link 
between nuclear activity and galaxy environment. 
However, at odds with that, the closest quasar pairs seem to inhabit poorer environments. 
Implications of present results and perspectives for future work are briefly discussed.
\end{abstract}




\section{Introduction}\label{theintro}
 \begin{table*}
 \begin{center}
 \footnotesize
 \caption{Quasar pair sample and properties.} 
 \label{sample}
 \begin{tabular}{@{}lcccccrrlll}
  \hline\hline
QP 	&QSO$_A$  			&z$_A$       & QSO$_B$  			&z$_B$ 	 &$\Delta\theta$ &$R_\bot$&$\Delta V_\parallel$&ref$_A$    &ref$_B$    &ref$_{QP}$\\
  	&(J2000)			&	     &(J2000)				&	 &[arcsec]  	 	 &[kpc]    &[km s$^{-1}$]             &            &  		&  	  \\   	  
 (a)    &(b)      			& (c)	     & (d)          			&(e)     &(f)            &(g)	   &(h)		       &(i)  	    &(j)  	&(k)      \\
  \hline
 &&&&&&&&&&\\
\ 1	&SDSS  J001103.18+005927.2	&   0.48646  & SDSS  J001103.48+010032.6 	& 0.48636  & 65	 & 394    & 20  $\pm$ 30  &Sc10	& P17	& S14	\\
\ 2 	&SDSS  J011757.99+002104.1	&   0.61127  & SDSS  J011758.83+002021.4  	& 0.61286  & 45	 & 300    &300 	$\pm$ 50  &Sc10	& Sc10	& H06	\\
\ 3	&SDSS  J022610.98+003503.9	&   0.42396  & SDSS  J022612.41+003402.2 	& 0.42383  & 66	 & 363    & 25  $\pm$ 40  &P17	& Sc10  & S14 	\\
\ 4	&SDSS  J023328.44$-$054604.4	&   0.49445  & SDSS  J023331.05$-$054550.9 	& 0.49394  & 41	 & 249    &100 	$\pm$ 20  &P17	& P17  	& S17	\\
\ 5	&SDSS  J074759.02+431805.3	&   0.50117  & SDSS  J074759.65+431811.4 	& 0.50175  & 8.9 & 56     &115 	$\pm$ 25  &Sc10	& Sc10	& H06	\\
\ 6	&SDSS  J074843.02+361258.7	&   0.65399  & SDSS  J074843.12+361219.4 	& 0.64959  & 39	 & 273    &795 	$\pm$ 35  &P17	& Sc10 	& S17	\\
\ 7	&SDSS  J082439.83+235720.3  	&   0.53526  &SDSS  J082440.61+235709.9 	& 0.53676  & 16	 & 94     & 290 $\pm$ 20  &Sc10	& Sc10	& H06	\\
\ 8	&SDSS  J084541.18+071050.3	&   0.53755  &SDSS  J084541.52+071152.3 	& 0.53516  & 62	 & 393    &470 	$\pm$ 50  &Sc10	& Sc10 	& H06	\\
\ 9	&SDSS  J085625.63+511137.0	&   0.54240  &SDSS  J085626.71+511117.8 	& 0.54316  & 23	 & 139    &150 	$\pm$ 20  &Sc10	& Sc10 	& H06	\\
10 	&SDSS  J093847.45+462328.2	&   0.57707  &SDSS  J093853.83+462310.8		& 0.57734  & 68	 & 448    & 50	$\pm$ 25  &P17	& P17	& S17	\\
11	&SDSS  J095137.00$-$004752.9	&   0.63395  &SDSS  J095139.39$-$004828.7 	& 0.63691  & 50	 & 346    &540 	$\pm$ 25  &P17	& P17	& S14	\\
12 	& 2QZ \ \ J115240.09$-$003032.8 &   0.55375  & SDSS J115240.52$-$003004.3 	& 0.55209  & 29	 & 188    &320 	$\pm$ 60  &C04 	& Sc10 	& H06	\\
13 	&SDSS  J115822.77+123518.5 	&   0.59572  & SDSS J115822.98+123520.3		& 0.59690  & 3.6 & 24     &220 	$\pm$ 60  &M08	& M08 	& M08	\\
14	&SDSS  J124031.42+111848.9	&   0.58480  & SDSS J124032.67+111959.2  	& 0.58404  & 73	 & 479    &145 	$\pm$ 60  &P17	& P17	& S17	\\
15	&SDSS  J124856.55+471827.7	&   0.43859  &SDSS  J124903.33+471906.0 	& 0.43861  & 79	 & 447    &  5  $\pm$ 15  &Sc10	& Sc10 	& F11	\\
16	&SDSS  J125454.86+084652.1  	&   0.43977  &SDSS  J125455.09+084653.9  	& 0.43969  & 3.8 & 22     &430	$\pm$ 70  &P17& Sc10 	& G10	\\
17 	&\ \ 2SLAQ J133350.41$-$003309.3&   0.60697  &\ \ 2SLAQ J133351.17$-$003248.3	& 0.61030  & 24	 & 160    & 620 $\pm$ 45  &C09 	& C09	& S17	\\
18 	&SDSS J141855.41+244108.9    	&   0.57305  & SDSS  J141855.53+244104.6 	& 0.57511  & 4.5 & 29     & 390 $\pm$ 30  &Sc10	& M08 	& M07	\\
19	&SDSS J155330.22+223010.2  	&   0.64127  & SDSS  J155330.55+223014.3 	& 0.64223  & 5.9 & 42     &175 	$\pm$ 15  &Sc10	& P17	& S14	\\
20 	&SDSS J164311.34+315618.4   	&   0.58653  &SDSS   J164311.38+315620.6	& 0.58636  & 2.3 & 15     & 30 	$\pm$ 30  &Sc10	& B99	& Mo99	\\
 &&&&&&&&&&\\
\hline
 \end{tabular}
\end{center}
\begin{tablenotes}
\item 
Notes:
(a) QSO pair (QP) identifier.
(b) and (d)  Names of  QSOs from SDSS, 2QZ and 2SLAQ surveys.
(c) and (e) Quasar redshifts derived from  [OIII]$\lambda$5007 \AA \  lines, see Sections \ref{thesample} and \ref{thespectra}.
Intrinsic wavelength-calibration uncertainties are added in quadrature to the line position 
errors.
(f) Angular separation between the two QSO pair components. 
(g) Proper traverse separation. 
(h) Radial velocity difference derived  from data in columns (c) and (e).
(i) and (j)  References  of the spectroscopic information used for the QSO pair search. 
(j) Reference of the first spectroscopical identification as a QSO pair.
\\
References:
B99: \cite{Brotherton1999}; 
C04:  \cite{Croom2004}, 2dF QSO Redshift Survey (2QZ);
C09: \cite{Croom2009}, 2dF SDSS LRG (luminous red galaxy) and QSO (2SLAQ) survey;
F11:  \cite{Farina2011};
G10:  \cite{Green2010};
H06: \cite{Hennawi2006}; 
Mo99:  \cite{Mortlock1999}; 
M07:  \cite{Myers2007};
M08:  \cite{Myers2008};
P17:  \cite{Paris2017}, SDSS Quasar Catalog  Data Release 12 (DR12Q); 
S14: \cite{Sandrinelli2014};
Sc10: \cite{Schneider2010}, SDSS Quasar Catalog Data Release 7 (DR7Q); 
S17: This work.
\end{tablenotes}
\end{table*}

Quasars (QSOs) are luminous and short-lived active galactic nuclei (AGNs).
They are associated with massive black holes (BHs) in the centre of 
galaxies powered by gas accretion and  they are revealed by their huge luminosity
\citep{Salpeter1964,Zeldovich1964,Lynden1969}.
To supply the enormous amount of gas  from kilo-parsec galaxy scale to the centre,  
different physical processes are invoked for luminous QSOs.
The most accredited models are those that involve major mergers  of similar-mass gas-rich 
 galaxies \citep[see e.g.][and references therein]{Dimatteo2005,Hopkins2008,Kormendy2013}. 
As an alternative,  models based on radiative instabilities  funnelling in the inner several kpc 
metal-rich stellar-remnant  recycled gas \citep[e.g][]{Ciotti2010}  have  been developed.
Less violent mechanisms, like minor mergers \citep[e.g.][]{Hopkins2008}, or secular processes
 unrelated with the merging phenomenon \citep[e.g.][]{Cisternas2011} 
as disk and  bar instabilities \citep[e.g.][]{Shlosman1989,Bournaud2011,Kocevski2012} 
 or stochastic events,  are also considered sufficient for fuelling less luminous AGNs
\citep[see e.g.][for a review]{Heckman2014}. 

Merger events are thought  to depend on the global properties of the 
galaxy environment \cite[e.g.][]{Kauffmann2000,Dimatteo2005}. 
 In a number of cases  galaxies hosting  QSOs are observed  in interacting systems or 
in   apparent merger products \citep[e.g.][]{Canalizo2001,Green2011,Kunert2011,Shields2012}.
Nevertheless, a significant enhancement of merger features in the QSO hosts with respect
 to those in inactive galaxies    was not observed
 \citep[e.g.][]{Dunlop2003,Mechtley2016,Villforth2017}. 
It was  also  expected that  QSOs should  preferably reside overly clustered regions
especially at high redshift \citep[e.g][]{Djorgovski1999,Volonteri2006}, 
given that mergers should be more frequent in denser environments.
However, not concordant  evidences of richer environments around  single QSOs in comparison to  
 inactive galaxies  were found, at both high \citep[e.g.][]{Morselli2014,Simpson2014} 
and intermediate-low redshifts \citep[e.g.][]{Serber2006,Padma2009,Shen2013,Zhang2013,
Karhunen2014,Krolewski2015,Jiang2016}. 
Evidences for a connection between environment, mergers and QSOs remain inconclusive. 
    
   Because of the short nuclear activity lifetime, only a small fraction of massive
 galaxies are shining as QSOs.
In spite of this, a significant  number of probable 
physical QSO pairs, i.e. QSO likely to be
mutually  gravitationally bound, belonging to the same cosmological structure,  were discovered on 
 sub-Mpc scale  \citep[e.g.][]{Hennawi2006,Myers2008,Hennawi2010,Sandrinelli2014}. 
The  presence of an enhanced  excess of these QSO pairs down to a scale of few tens kpc
enforced the  scenario of  tidal interactions in gas-driven mergers leading to  the  mutual triggering  
of  the nuclear phase in both QSOs 
\citep[e.g.][]{Djorgovski1991,Kochanek1999,Mortlock1999,Myers2008,Foreman2009}.
In such a case, a comparable small-scale enhanced environment is expected.
The simultaneous presence of two close QSOs  could also  be  explained as a statistically
 predictable consequence of  group-scale environments, in which  small scale  galaxy 
 overdensities make mergers 
 more likely to occur \citep[e.g.][]{Hopkins2008},
or as a manifestation of the clustering properties of the dark matter haloes hosting the 
QSOs \citep[e.g.][]{Dimatteo2005,Richardson2012}. 
In the sketched pictures, QSO pairs constitute special cases to probe 
 the quasar phenomenon.
They could pose constraints on how, and to what extent, the galaxy environment is connected to 
 the QSO activation, or clear up if quasar phase is  stochastic process that 
every luminous galaxy in a typical galaxy cluster can experience.
 In the latter case, QSO pairs simply could  derive from two galaxies simultaneously active 
 for a period. 

The QSO pairs environment  has been investigated  only 
in a few papers  due to the lack of large samples and suitable data,
resulting on average in a poor enhanced environment of galaxies at small scale, although 
signatures of  galaxy clusters were detected in some cases.
No enhancement in the galaxy density around a luminous z=4.25 QSO  pair was found by \cite{Fukugita2004}.
\cite{Boris2007} explored the photometrical  properties of 
 four fields around QSO pairs at z$\lesssim$1 with separations of about 1 Mpc, 
leading to mixed  results.
In a study on six  physical QSO pairs at z$\leqslant$0.8 drawn from the  Sloan Digital Sky Survey (SDSS),
\cite{Farina2011} reported  one case of  pair in a significant  overdense group of galaxies 
and  dynamical evidences of mass  exceeding that observed around  other targets.
\cite{Green2011}, searching for signs  of galaxy clusters associated with seven close QSO
pairs at z$\sim$1, did not detect  such an evidence. 
 \cite{Sandrinelli2014} investigated the environment 
of 14 physical SDSS QSO pairs at z$<0.85$ and found that they are harboured on average in  regions 
of modest galaxy overdensity   extending up to $\sim$0.5 Mpc,
suggesting that the rare activation of two QSOs with small physical separation does
not require an extraordinary environment.
Recently, \cite{Onoue2017},   applying  less stringent 
constraints than ours (see Section \ref{thesample}) in defining QSO pairs,
investigated the overdensities around  associations of  two QSOs    
extracted from the SDSS-III BOSS Survey in the DR12Q catalog \citep{Paris2017} 
with cluster-scale separation, as possible  massive proto-clusters tracers.
Focusing on the 33 pairs at $z\sim1$,  they reported evidence
 of  enhanced environments  in $\sim$ 20 \% of cases.
At higher z (3$<$z$<$4) they also detected two QSO pairs 
 in a significant overdense environment.
 Rare cases of  associations with more than two QSOs  have been
observed at 1.5 $\lesssim$ z $\lesssim$ 2   by \cite{Djorgovski2007}, 
 \cite{Farina2013},  and  \cite{Hennawi2015} 
 in very different environments (from poor to substantially overdense).
In general, in previous works no exhaustive  conclusion has been yielded on the subject.

To further investigate a possible link between the environment and  the quasar activity, 
in this work we use SDSS images to analyse the environment properties
at  $\sim$ 1 Mpc-scale  of a homogeneous sample of 
20 reliable QSO physical pairs   at 0.4 $<$ z $\lesssim$ 0.65, selected from all the available QSO 
data sets (Section \ref{thesample}). 
The redshift range was chosen to  better  reveal possible environmental galaxy excess
on the SDSS images.
We complement  SDSS imaging (Subsection \ref{theim}) and  spectroscopic 
studies  with  high quality optical spectroscopy gathered at Gran Telescopio
 CANARIAS (GTC) in La Palma (Subsection \ref{thespectra}).
As comparison sample, a purpose-built sample of 200  isolated QSOs
 drawn from SDSS archives,  matching   in redshift and luminosity with the QSO in pairs
 was selected (Section \ref{isolQ}).
For the QSO pairs we perform a detailed analysis of the clustering (Sections \ref{theenv} and \ref{thenng}) 
of galaxies and  compare the results with those derived from the control sample of isolated QSOs. 
 The results and  their implications are discussed (Section \ref{thediscussion}).
 
In this work we assume a concordant  cosmology:  
H$_0=70$ km  s$^{-1}$     Mpc$^{-1}$, $\Omega_m =0.30$ and $\Omega_\Lambda=0.70$.\\

 \begin{figure}
 \includegraphics[width=1\columnwidth]{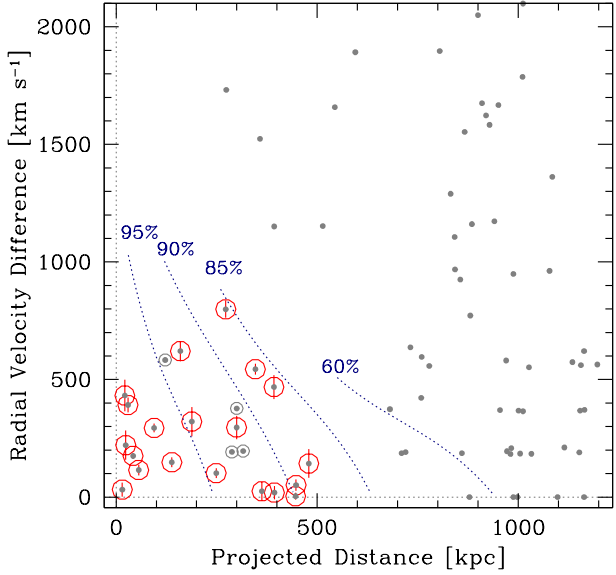} 
  \caption{\label{pddv}
   Distribution   in  the     $R_{\bot}-\Delta V_{\parallel}$ plane  of QSO   pairs  at 
   0.4 $<$ z $<$ 0.66 (grey dots).
The percentage of the  QSO pairs expected to be physical systems ranges  between  the reported values. 
The pairs investigated in this work are  marked by red large circles.
Excluded candidates  (see the text)  are surrounded by grey small circles.
 }
\end{figure}

\section{The sample}\label{thesample}

We searched for  physical QSO pair candidates following the procedure 
outlined in  \cite{Sandrinelli2014}. 
The starting data set is based on  the spectroscopic QSOs collected  in 
The Million Quasars (MILLIQUAS) Catalog, Versions 3.9-4.3  \citep{Flesch2015}, 
containing  more than 420,000 QSOs from literature. 

 We expected to reach  apparent magnitudes of $\sim$ 22  in the SDSS i-band images  
 (see Section \ref{theenv}). At these magnitudes, we wished to detect environmental 
 galaxies up to a limit  $\gtrsim$ M*  \citep[M*=$-$21.70 mag,][]{Montero2009} 
  around  the highest  redshift QSO in our sample. We also wanted to limit  the number of detectable 
 background galaxies, which could overshadow possible local overdensity effects at low redshifts.
 With these aims, we  restricted the QSO pair sample search to $\sim$ 21,500 QSOs  at 
 0.4 $ < $ z $\lesssim$ 0.65. 
In addition, at these redshifts  QSOs  show the  [OIII]$\lambda$5007 line well 
inside the optical spectral range, allowing us to improve the accuracy in
 measuring their systemic radial velocity
 for a more reliable selection  of probable bound QSO pairs
(see below, and also see Section \ref{thespectra}).

For each association of two QSOs, constructed by coupling  each QSO  with all the others, 
we evaluate projected separations $R_{\bot}$ and differences of  radial 
velocities $\Delta V_\parallel$ between the two components.
To assemble a  population of binary QSOs likely to be gravitationally bound,
in each bi-dimensional bin of the  $R_{\bot}$-$\Delta V_\parallel$ plane
(Figure \ref{pddv}) the excess of  observed  QSO pairs with respect  to those 
obtained randomly permuting the redshifts  \citep[e.g.][]{Zhdanov2001} is estimated. 
This allows us to trace the loci in $R_{\bot}$-$\Delta V_\parallel$ plane
where the probability for a QSO pair of being not a chance superposition 
is higher than a fixed constant value $\xi$.
For our search we choose $\xi$$>$85\%.
  
 After a first selection based on  redshifts in literature,
 24 \ QSO physical pairs candidates were found.
Four of them were discharged for dubious classification of one component after 
visual inspection of the spectra drawn from SDSS\footnote{The discarded QSO physical pair candidates are
 J015628.41+174957.6 \citep[SDSS DR10,][]{Ahn2014}, 
 J091442.32+000637.1 \citep[SDSS DR3,][]{Abazajian2009},
J103834.84+595725.7 and  J121336.12+463355.8 \citep[SDSS DR12,][]{Alam2015}.
Their spectra are characterized by  flat-shape continuum, absence of broad 
lines and  suppressed MgII  emission line.  
 We note that none of these objects is recovered in SDSS DR7Q \citep{Schneider2010} 
 and DR12Q  \citep{Paris2017}. 
}.
A dedicated complementary spectroscopical program was undertaken at the 
Gran Telescopio Canarias (GTC, Subsection \ref{thespectra})
for a number of the remaining pairs, not investigated in \cite{Sandrinelli2014}.
 We re-evaluate the positions of  all the selected  QSO pairs in the $R_{\bot}$-$\Delta V_\parallel$ 
 plane, by measuring $\Delta V_\parallel$  from the narrow forbidden[OIII]$\lambda$5007 
 emission line, which arises  from  regions  where  gas is
 predominantly  orbiting in the  host galaxy potential  and therefore is a  good z estimator
\citep[e.g.][]{Hewett2010,Liu2014}.
The wavelength of the line was measured with the procedure described in \cite{Farina2011}.

All  20  pairs (z$_{ave}$=0.55) were confirmed in the final sample of physical QSO pairs candidates  
(see Section \ref{thespectra}), of which 5 are new,  not discussed before.
The sample is  given in Figure \ref{pddv} and Table  \ref{sample},  where  $\Delta V_\parallel$ from [O III]${\lambda5007}$ 
measurements are reported.
It contains  nine  pairs  already studied in  \cite{Sandrinelli2014}. 
Seven pairs have   $R_{\bot} <$ 100 kpc.
The sample is largely  radio-quite 
dominated\footnote{Data from VLA FIRST  Survey (\texttt{http://sundog.stsci.edu})
and NRAO VLA Sky Survey (\texttt{http://www.cv.nrao.edu/nvss})
catalogues.}, 
with only one QSO detected as radio-loud (J164311.34+315618.4, see also Appendix \ref{appA}).

 \begin{table}
 \begin{center}
 \caption{Photometrical properties of the subsample of newly investigated QSO pairs.} 
 \label{photom}
 \begin{tabular}{@{}rccllcc}
  \hline\hline
QP 			& i$_A$ 	&M$_A(i)$  	 &i$_B$    &M$_B(i)$	&$m_{i,50\%}$ 	&$M_{i,50\%}$\\
  	  	 	& [mag]  	& [mag]  		 &[mag]     	&[mag]       	&[mag]  	    		&[mag]       	        \\   	  
 (a)      		&(b)      	& (c)	      		&   (d)        &(e)			& (f)          			&(g)	  	        \\
  \hline
  &&&&&&\\
2 	&19.99    &$-$22.87	&18.25    &$-$24.62 & 22.2  	&$-$20.68     \\ 	  	
4	&20.31    &$-$21.97	&18.45    &$-$23.83 & 22.0  	&$-$20.33     \\	
6 	&19.97    &$-$23.14 	&20.19    &$-$22.91 & 22.0  	&$-$21.09     \\	
10	&19.30    &$-$23.35	&19.97    &$-$22.70 & 21.9  	&$-$20.83     \\ 	  	
12	&20.04    &$-$22.53 	&18.84    &$-$23.72 & 22.1  	&$-$20.52     \\	
13	&19.49    &$-$23.30 	&19.77    &$-$23.00 & 21.8  	&$-$21.03     \\	
14	&19.99    &$-$22.76 	&20.18    &$-$22.55 & 21.9        &$-$20.87     \\	
16	&19.29    &$-$22.69	&17.06    &$-$24.91 & 21.9  	&$-$20.14     \\	
17     &19.96    &$-$22.90	&20.39    &$-$22.47 & 21.9  	&$-$20.97     \\	
18	&18.94    &$-$23.72	&19.89    &$-$22.77 & 22.0       &$-$20.72    \\	
20	&18.80    &$-$23.95	&19.51    &$-$23.23 & 22.1  	&$-$20.68     \\	
 &&&&&&\\
 \hline
 \end{tabular}
\end{center}
\begin{tablenotes}
\item 
Notes: 
(a) QSO pair identifier. 
(b) and (d)  SDSS  i-band apparent magnitude (psfmag) of QSO A and B, respectively; 
(c) and  (e) Extinction and k-corrected absolute  magnitude of QSO A and B, respectively. 
(f) and (g)  Apparent SDSS magnitude threshold and correspondent  absolute
magnitude, see Section \ref{theenv}.
\end{tablenotes}
\end{table}

 \begin{table*}
 \begin{center}
\caption{\label{journal}
Journal of GTC observations.
}
\begin{tabular}{@{}rcccrlcll}
\hline\hline
QP     	&Date 			&Seeing	&Slit	&t$\rm_{exp}$	&Grism	& SNR$_{\rm(A)}$& SNR$_{\rm(B)}$	&SDSS	\\
		&				&[arcsec]	&[arcsec]&[s]			&	    	&			&		 	&				\\
	(a)	&	(b)			&(c)		&(d)	&(e)			&(f)		&(g)	    		&(h)			&(i)		 		\\
\hline
&&&&&&&&\\
	2	 & 2015 December 	28	&1.5		&1.0	&900 	  	&R2500I	&12			&25					&A, B	\\
 	12	 & 2016 March  	19	&0.9		&1.0	&980   		&R2500I	&10			&16					&B     	\\
 	13	 & 2015 April 		10	&1.4		&1.2	&800	 	&R2500I	&20			&14					&        	\\
	14	 & 2015 May 		10	&1.0		&1.2	&900 		&R2500I	&14			&10					&A, B	\\
 	16      & 2016 February  	06     &1.3		&1.0	&900 	 	&R2500R	&31			&70					&A, B 	\\
	17	 & 2016 March 		26	&1.0  	&1.0	&900   		&R2500I	&10			&8					&        	\\
	18	 & 2015 May 		10	&0.9		&1.2	&1000    		&R2500I	&25			&10					&A      	\\
	20	 & 2015 April  		10	&1.3		&1.2	&500  		&R2500I	&26			&15					&A      	\\
&&&&&&&&\\
\hline
\end{tabular}
\end{center}
\begin{tablenotes}
\item
Notes: 
(a) QSO pair  identifier. 
(b) Date of observations; 
(c) Mean seeing during the observations;
(d)  Slit width;
(e) Exposure time; 
(f) Grism;
(g) and (h) Average 
signal-to-noise ratio of the spectrum of  QSO A and QSO B, respectively; 
(i)  QSO spectroscopically observed by SDSS.
\end{tablenotes}
\vspace{1cm} 
\end{table*}

\section{Observations}
\subsection{SDSS Imaging}\label{theim}

Calibrated, sky-subtracted i-band images of the QSO fields were retrieved from SDSS archives. 
At the average redshift of the QSO data set, this corresponds to observe in 
the SDSS g filter at rest frame.
Images have an exposure time of 54 s and a mean seeing of 1.04 arcsec.
 We drew photometrical information from  SDSS DR12 catalogues,
  where objects classifications are based on the difference between the cmodelMag
(composite de Vaucouleurs and exponential model) and the point spread function (psf)  magnitudes.
 We obtained position and i-band photometry  of  QSOs  and of  all primary objects photometrically 
classified as galaxies (\texttt{type = 3}) in the fields.  
The  photometrical properties of the QSO pair sample given
 in Table  \ref{photom} complement those already reported in  \cite{Sandrinelli2014}.
Apparent magnitudes (psfmag) are taken from SDSS data base.  
Absolute magnitudes M$_i$ are extinction corrected on the basis of SDSS values
 following \cite{Schlegel1998}.
\textit{K}-corrections are also applied  by adopting the 
template spectra of  \cite{Francis2001} and \cite{Mannucci2001} for QSOs and galaxies, respectively,
  and  the i-band filter response.
The mean i-band absolute dereddened magnitude  for the full sample
is  M$_{ave}$(i) = $-$23.16 $\pm$ 0.12 mag, where the uncertainty is the standard error of the mean.

\subsection{Optical spectroscopy}\label{thespectra}

Among the selected  QSO pairs in Table \ref{sample}, five lack of spectra in 
SDSS archives for at least one QSO member (see Table \ref{journal}).
Three of them  have small component separations 
 (2$\arcsec$$<\Delta \theta<$ 10$\arcsec$)
and they  are of special interest. 
 Although in general considered as highly probable close binaries rather than 
 lensed sources (see Table \ref{sample} and Appendix \ref{appA} for more details),
 the lensing hypothesis was not excluded for some cases   
 \citep[QSO pairs QP13 and  QP18,][]{Myers2008,Myers2007},
deserving higher resolution spectroscopy.
Another QSO pair is newly discovered  (QP17).
Both its QSOs were independently   observed in the 2SLAQ Survey \citep{Croom2009}.  
For the remaining  pairs (QP2, QP14, QP16, 
of which the last one is another close pair),  
we were interested in obtaining  reliable measures of the  [OIII]$\lambda$5007 region. 

\begin{figure}
\includegraphics[width=88mm]{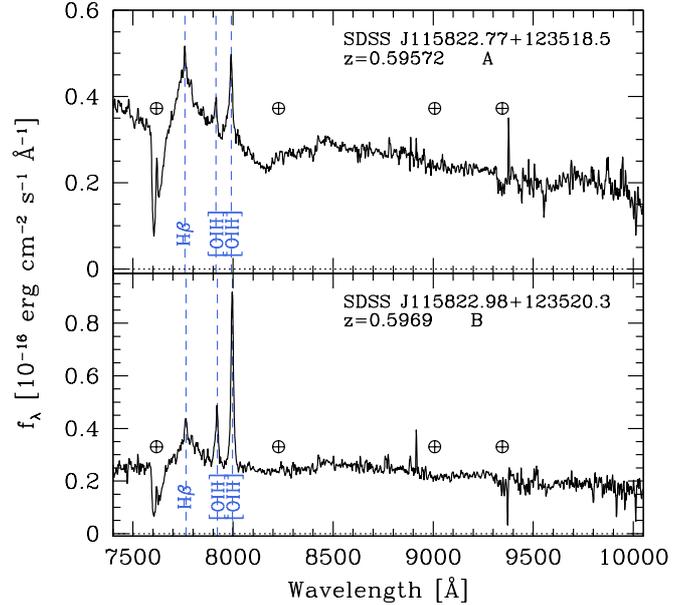}
  \caption{\label{gtcex}
GTC spectra of the QSO pair QP13 (see also Appendix \ref{appA}).
The two components have a projected separation of 3.6 arcsec,
corresponding to 24 kpc at the redshift of the pair,
and a difference of radial velocity of  (220 $\pm$ 60) km s$^{-1}$ (Table \ref{sample}).
The most prominent  emission features  are  marked. 
The main telluric bands are indicated by $\oplus.$
The comparison between the two spectra allows us to 
exclude  that the QSO pair is a gravitationally lensed image.
}
\end{figure}

For  these QSO pairs we secured optical spectra at  the 10.4m GTC, located in 
Roque de los Muchachos, La Palma, using the OSIRIS spectrograph \citep[][]{Cepa2003}.
Intermediate resolution (R $\sim$2200) observations were performed using
 R2500I or R2500R grisms  in order to detect the  [OIII]$\lambda$5007 \AA \  line region.
 For each QSO pair the slit was oriented  to simultaneously secure the spectra  
 of both  objects, and three individual spectra were obtained.
We reduced  data by using the  standard IRAF recipes
and applied flux corrections using SDSS photometry.
Journal of GTC observation is given in Table \ref{journal}.
In Figure \ref{gtcex} we show as example the GTC  spectra of the QSO pairs QP13. 
The other ones are illustrated in Figure \ref{gtc1} in Appendix \ref{appA},
 where  notes on some individual paired targets are also reported.
Detailed comparisons between the two GTC optical  spectra of each pair allow us to
confirm that all our targets are physical QSO pair candidates.  
In particular, the QSOs  with small angular separation,
are paired QSOs rather than lensed images,
see Appendix \ref{appA}.

\section{A comparison sample of isolated quasars}\label{isolQ}

 \begin{figure}
\includegraphics[width=1\columnwidth]{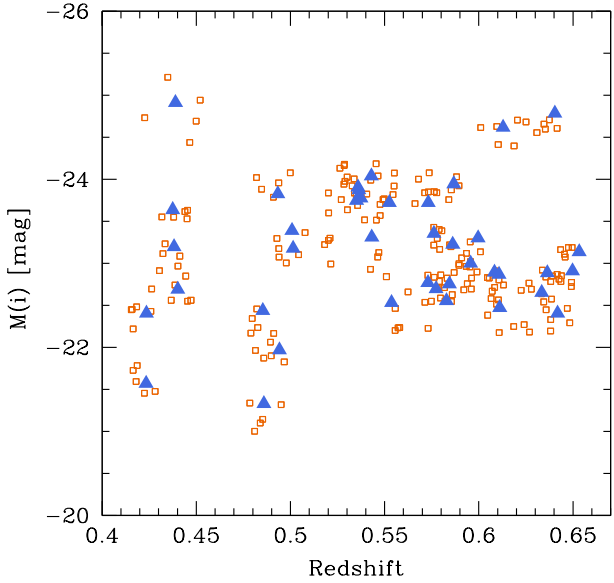}  
  \caption{\label{Mired}
 The QSO samples in the M(i)-redshift plane. 
Blue triangles are the  QSOs in pairs. The 200 isolated QSOs
of the comparison sample, matched in redshift and luminosity with QSOs in pairs, 
are  reported as  orange open squares.  
}
\end{figure}

We aim to probe  whether the clustering of galaxies in   $\sim$1 Mpc-scales region hosting QSOs 
is sensitive to the presence of QSO pairs. 
In particular, we  compare the average overdensity  distribution of galaxies 
around  individual  QSOs in pairs with that  of 200 isolated QSOs, 
 which are well matched with the paired QSOs in terms of redshift and luminosity. 
The sample was assembled  from SDSS DR12 
Catalog\footnote{\texttt{http://www.sdss.org/dr12/algorithms/boss-dr12-quasar-catalog}} 
of spectroscopical confirmed QSOs, following Subsection \ref{theim}
 for photometrical  measurements.
 We randomly extracted ten isolated sources for each QSO pair,
drawing five objects among those  at distances |$\Delta$\textit{z}| $<$ 0.02 and |$\Delta$\textit{M}$_i|<$ 0.35 mag 
 from each QSO in pairs
in the redshift-luminosity plane  (see Figure \ref{Mired}).
Before confirming objects in the sample,
optical SDSS spectra were visually checked  to secure  identification,  
to reliably estimate the emission redshift, and to identify peculiar 
 features.
The two samples of paired and  isolated QSOs also well match in terms of
 distributions of host galaxy luminosity, both ranging from $-$21 mag to $-$25.5 mag 
around the mean value $ <$M(r)$_{host}$$>$$\sim$ $-$23.3 mag.  
Details on the nuclear/host luminosity decomposition
of the QSO images  are given in Appendix \ref{appC}.

\section{ Galaxy environment of the quasar pairs }\label{theenv}

In order to characterize the environment around QSOs in  pairs and around isolated QSOs, we estimated the
 surface density  of galaxies with respect to the distance from the targets using the SDSS objects classified as galaxies, see Section \ref{theim}.
The midpoint of each pair was used to evaluate the background galaxy density at angular
 distances  between 7 and 15 arcmin, corresponding to a  projected distance between $\sim$ 2.5 and  $\sim$ 5 Mpc for the nearest pair. 
The environment around QSOs  is then evaluated  around the position of each individual QSO.

\begin{figure}
\centering
\includegraphics[width=1\columnwidth]{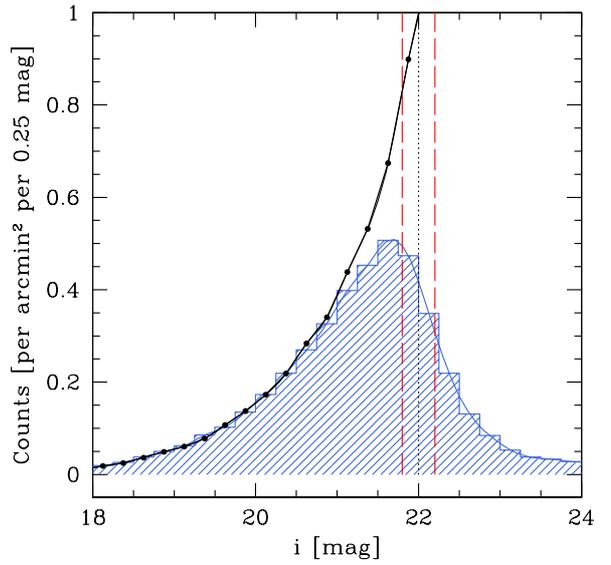}
 \caption{ 
 \label{galhist}  
Mean  i-band magnitude   distribution of the galaxies in the background regions around the QSO pairs
(see the text).  Black solid line represents the expected distribution of galaxies \citep[][]{Capak2007}. 
Dotted line marks the 50\% completeness limit $m_{i,50\%}$, see Table \ref{photom}.
Red large-dashed lines indicate the minimum and the maximum magnitude thresholds 
 in the sample. 
}
\end{figure}

\begin{figure*}
 \includegraphics[width=1\columnwidth]{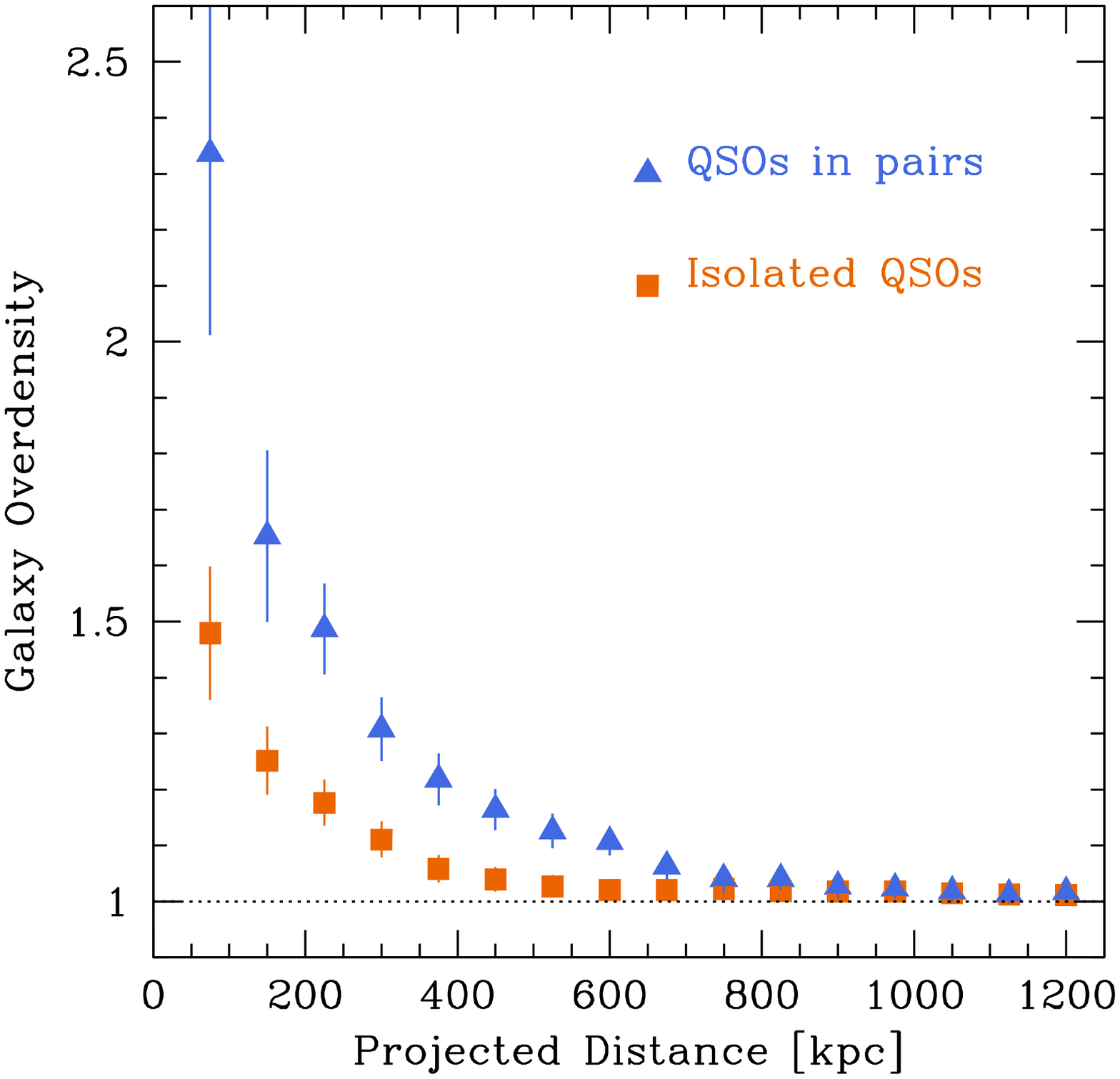}
\hspace{-0.5cm}
\includegraphics[width=1\columnwidth]{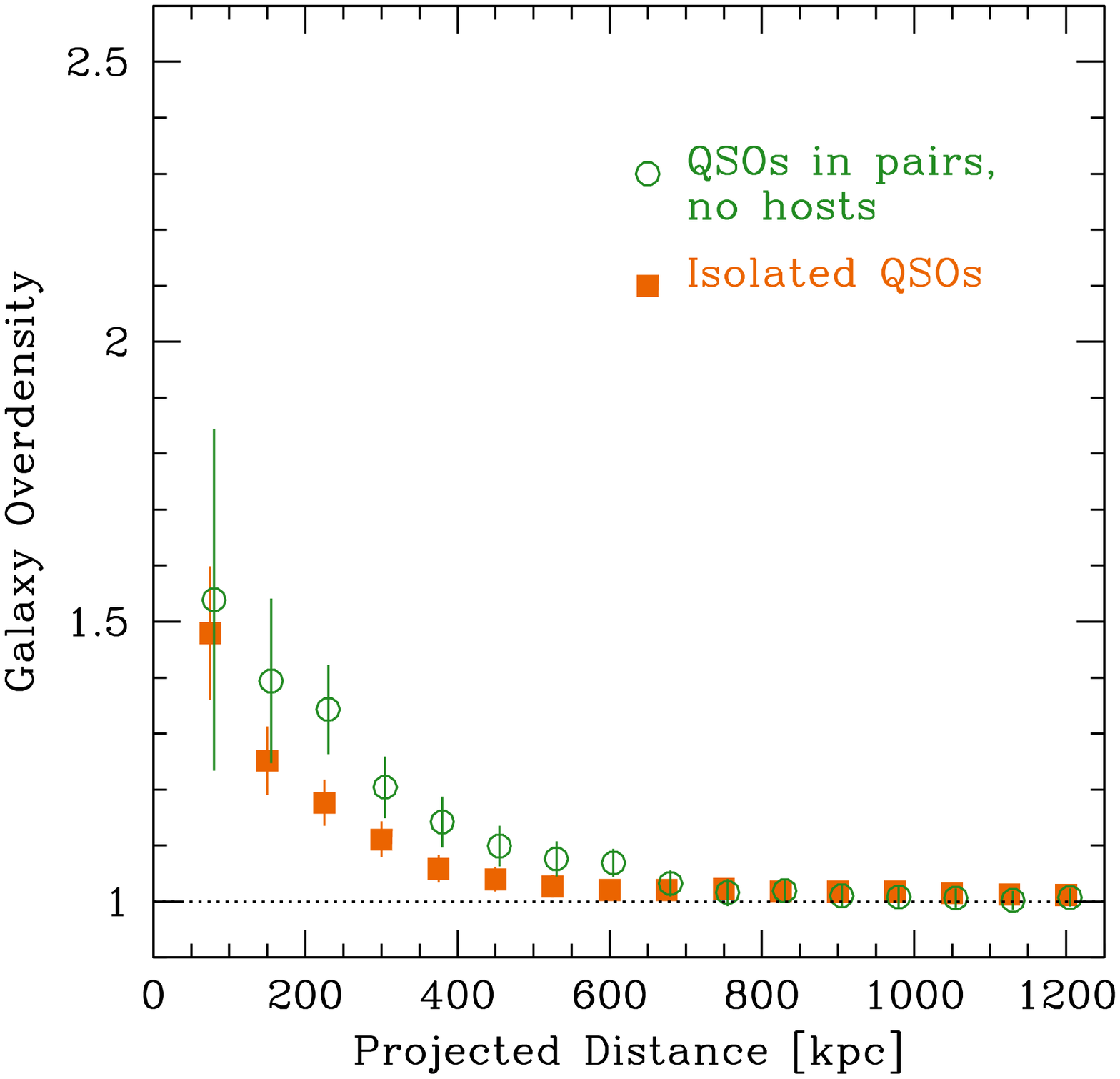}
 \caption{\label{Gnohost1}
 Mean cumulative galaxy overdensities  around  QSOs in pairs 
as a function of the projected distance from the QSO.
The cases of inclusion and exclusion  of the host galaxy of the companion QSO 
are reported, respectively, in the \textit{left panel}  (blue triangles), and in the \textit{right  panel} 
(green circles). Mean cumulative galaxy overdensity  around  200 isolated QSOs of the 
comparison sample, matching with the QSO pairs  in luminosities and redshifts, is also 
plotted (orange squares).
The uncertainties are the standard errors of the mean evaluated in each  bin.
  }
\end{figure*}

\begin{table*}
\begin{center}
\caption{
Statistics of galaxies in the QSO pair  environments.
}
\label{stat}
\small
\begin{tabular}{rcccccccccccc}
\hline\hline
QP  	&n$_{bg}$      	     &n$_{bg}$      	   &N$_{0.25}$ &N$_{0.25}$&N$_I$	&N$_{bg,I}$&$\mathcal{N}_{0.25}$ 	&$\mathcal{N}_{0.25}$	&$G_{0.25}$ 	&G$_{0.25}$	&G$_{0.5}$  &G$_{0.5} $	\\	 
	&[arcmin$^{-2}$]   &[Mpc$^{-2}$]	    &A			&B		&		&		   &A					&B					&A			&B			&A		    &B			\\
(a)	&(b)			     &(c)			    &(d)			&(e)		&(f)		&(g)	 &(h)	 &(i)	&(j)	&(k)	&(l)	&(m)		\\

\hline
&&&&&&&&&&&&\\
1 	& 3.6  $ \pm$  0.1  &  27.4  $ \pm$  0.8   & 7 &    6  &     0 &   0.6  &    7.0   	& 6.0		&1.30  $ \pm$  0.04 &  1.11 $ \pm$  0.02 &  1.21 $ \pm$  0.03 &  0.70  $ \pm$  0.02	\\
2 	& 3.9  $ \pm$  0.3  &  23.7  $ \pm$  1.7   & 8 &    7  &     3 &   1.3  &   7.2	& 6.2		&1.54  $ \pm$  0.13 &  1.32 $ \pm$  0.08 &  1.17 $ \pm$  0.09 &  1.01  $ \pm$  0.08	 	\\
3 	& 4.0  $ \pm$  0.3  &  36.2  $ \pm$  2.3   & 9 &    12 &     3 &   1.2  &  8.1	& 11.1	&1.14  $ \pm$  0.08 &  1.56 $ \pm$  0.09 &  1.17 $ \pm$  0.08 &  1.31  $ \pm$  0.09	 		\\
4	& 3.7  $ \pm$  0.2  &  27.9  $ \pm$  1.2   & 9 &    9  &     3 &   2.2  &   8.6	& 8.6		&1.57  $ \pm$  0.06 &  1.57 $ \pm$  0.05 &  1.17 $ \pm$  0.05 &  1.12  $ \pm$  0.05	 	\\
5 	& 3.2  $ \pm$  0.2  &  24.0  $ \pm$  1.3   & 8 &    9  &     7 &   4.0  &   6.5	& 7.5		&1.38  $ \pm$  0.08 &  1.59 $ \pm$  0.06 &  1.07 $ \pm$  0.06 &  1.07  $ \pm$  0.06	 	\\
6 	& 3.3  $ \pm$  0.2  &  19.3  $ \pm$  1.1   & 9 &    5  &     2 &   1.3  &   8.7	& 4.7		&2.28  $ \pm$  0.13 &  1.23 $ \pm$  0.09 &  1.12 $ \pm$  0.07 &  1.52  $ \pm$  0.09	 	\\
7 	& 3.6  $ \pm$  0.3  &  24.9  $ \pm$  1.7   & 11&   7   &     6 &   3.7  &  9.9	& 5.9		&2.02  $  \pm$  0.14 & 1.20  $ \pm$  0.10 & 1.34  $ \pm$  0.10 & 1.34  $ \pm$  0.10	 	\\
8 	& 3.5  $ \pm$  0.2  &  24.4  $ \pm$  1.2   & 5 &    7  &     1 &   0.5  &   4.8	&  6.8	&0.99  $ \pm$  0.05 &  1.41 $ \pm$  0.05 &  0.88 $ \pm$  0.05 &  0.93  $ \pm$  0.05	 		\\
9 	& 3.3  $ \pm$  0.1  &  22.9  $ \pm$  0.9   & 4 &    4  &     2 &   2.9  &   4.0      &  4.0	 &0.89 $ \pm$  0.03 &  0.89 $ \pm$  0.03 &  0.84 $ \pm$  0.03 &  0.95  $ \pm$  0.03	 	\\
10 	& 3.3  $ \pm$  0.1  &  21.5  $ \pm$  0.7   & 7 &    4  &     1 &   0.2  &    6.6	&  3.6	&1.56   $ \pm$  0.05 & 0.85  $ \pm$  0.03 & 1.41  $ \pm$  0.04 & 0.88  $ \pm$  0.03	 	\\
11	& 3.1 $ \pm$  0.2  &  18.2  $ \pm$  0.9  & 7 &    4  &     2 &   0.7  &    6.3	&  3.3	&1.78   $ \pm$  0.09 & 0.94  $ \pm$  0.07 & 1.20  $ \pm$  0.06 & 1.34  $ \pm$  0.07	 		\\
12 	& 4.3  $ \pm$  0.2  &  29.0  $ \pm$  1.2   & 14&   16 &    8 &   3.0  &  11.5	& 13.5	&2.02   $ \pm$  0.10 & 2.38  $ \pm$  0.07 & 1.30  $ \pm$  0.06 & 1.44  $ \pm$  0.07	 		\\
13 	& 2.9  $ \pm$  0.2  &  17.9  $ \pm$  1.4   & 8 &    6  &     4 &   3.3  &   7.7	& 5.7		&2.18   $ \pm$  0.15 & 1.61  $ \pm$  0.13 & 1.52  $ \pm$  0.14 & 1.37  $ \pm$  0.13	 	\\
14	& 3.2  $ \pm$  0.3  &  20.7  $ \pm$  1.7   & 4 &    7  &     0 &   0.05  &    4.0	& 7.0  	&0.98   $ \pm$  0.08 & 1.72  $ \pm$  0.08 & 1.17  $ \pm$  0.09 & 1.04  $ \pm$  0.08	 	\\
15	& 3.9  $ \pm$  0.1  &  33.4  $ \pm$  0.9   & 9 &    9  &     1 &   0.3  &    8.6	&  8.6	&1.32   $ \pm$  0.04 & 1.32  $ \pm$  0.03 & 1.20  $ \pm$  0.03 & 1.12  $ \pm$  0.03	 	\\
16	& 3.2  $ \pm$  0.2  &  27.8  $ \pm$  1.9   & 8 &    7  &     6 &   5.2  &   7.6	& 6.6		&1.39   $ \pm$  0.09 & 1.21  $ \pm$  0.06 & 0.96  $ \pm$  0.06 & 0.96  $ \pm$  0.06	 	\\
17 	&2.9 $ \pm$  0.1  &  17.6  $ \pm$  0.6   & 8  &  4  &     2 &   2.1  &   8.0	& 4.0  	&2.32   $ \pm$  0.06 & 1.16  $ \pm$  0.04 & 1.09  $ \pm$  0.03 & 1.31  $ \pm$  0.04	 		\\
18 	& 3.3  $ \pm$  0.1  &  21.4  $ \pm$  0.6    & 2 &   2   &    1 &   3.9  &   2.0	& 2.0  	&0.48   $ \pm$  0.01 & 0.48  $ \pm$  0.03 & 1.01  $ \pm$  0.03 & 0.89  $ \pm$  0.03	 		\\
19	& 3.1  $ \pm$  0.2  &  17.9  $ \pm$  1.2    & 11&  11  &   9 &   3.1  &  8.1	& 8.1		&2.29   $ \pm$  0.18 & 2.29  $ \pm$  0.11 & 1.43  $ \pm$  0.11 & 1.36  $ \pm$  0.11	 	\\
20 	& 3.4  $ \pm$  0.1  &  21.9  $ \pm$  0.7    & 7  &  7   &     6 &   4.1  &  6.1	& 6.1		&1.41   $ \pm$  0.04 & 1.41  $ \pm$  0.02 & 0.87  $ \pm$  0.02 & 0.87  $ \pm$  0.02	 	\\
&&&&&&&&&&&&\\
\hline
\end{tabular}
\end{center}
\small
\begin{tablenotes}
\item
Notes: 
The target host galaxy is excluded from the counts and companion host is included. Errors are calculated taking into account only 
 the average background  surface density uncertainties. 
 (a)  QSO pair identifier.
(b) and (c)   Surface density of galaxies in the background  with
 i $<m_{i.50\%}$ in arcmin$^{-2}$ and Mpc$^{-2}$, respectively.
 (d) and (e) Number of galaxies within 250 kpc from the QSO A and B, respectively. 
(f)  Number of  galaxies in the  overlapping region.
(g) Number of expected background galaxies 
in the overlapping region, see the text. 
 (h) and (i) Number of galaxies within 250 kpc from each QSO after subdivision of the excess of 
 galaxies in common. 
  (j)  and (k) Galaxy  overdensity  within 250 
kpc from the QSO A and B, 
  corrected for the superposition of the companion environment.	
(l) and (m) The same  as in columns (j) and (k) within 500 kpc.
\end{tablenotes}
\end{table*}

To take into account the completeness of the SDSS galaxy catalogues,
galaxy surface densities are estimated by counting galaxies brighter than a magnitude limit.
This is fixed for each field at the  magnitude  $m_{i,50\%}$, where the differential 
magnitude distribution of the observed  galaxies in the background region 
drops to 50\% of that estimated by
the  deep galaxy survey in \cite{Capak2007}\footnote{Durham University Cosmology Group, 
references and data in \texttt{http://astro.dur.ac.uk/$\sim$nm/pubhtml/counts/counts.html}}.
In our SDSS i-band images the  threshold magnitudes  are 
distributed  around the mean value $m_{i,50\%,ave}$ =  22.0 $\pm$ 0.02 mag (see Table \ref{photom}).
At these thresholds we can observe galaxies brighter than $\sim$M*+2 at z=0.4, 
$\sim$M*+1  at z=0.5 and $\sim$M* at z=0.65.
The   mean  distribution  of  galaxy magnitudes obtained from the background regions 
 of the QSO pair sample is shown in Figure \ref{galhist}. 
We estimate the background surface density  $n_{bg}$ of galaxies  by computing the
median of the galaxy surface density observed  in  1 arcmin-width annuli in the background area.  
The local cumulative\footnote{Since each aperture includes the inner ones, 
we refer to quantities derived from this approach  as \textit{cumulative}, see \cite{Karhunen2014}.} 
surface density  $n(r)$ around each QSO  is evaluated 
inside circular apertures of radius $r$, with $r$ ranging  by steps of 75 kpc and  covering the inner 2.5 Mpc. 
The target QSO host galaxy is excluded. 
The cumulative overdensity  profile of galaxies is then defined  as   $n(r)/n_{bg}$.
 In the case of QSO pairs,  to take into account the contribution of  the two sources, for each 
 radius $n(r)$ is corrected by subdividing  in equal number to each QSOs the excess 
of galaxies (over the background) in the region where the apertures overlap 
\citep[see][]{Sandrinelli2014}. Hereafter, we refer to the overdensity as $G(r)$.
For the case of QSOs in pairs, details of  counts of   individual QSOs contribution to the pair 
environment are reported in Table \ref{stat}  for radius apertures  of 250 kpc, and results are
 also given in the case of 500 kpc radius.

Because of the low statistics, the $G(r)$ distributions for each source appear rather noisy 
 and significant differences among the various QSO are apparent. 
 Therefore we have concentrated on the averaged cumulative profile evaluated 
 of the entire sample  of the 40 QSO belonging to pairs.

 Our main result is that the galaxy overdensity of  QSOs in pairs is clearly larger
  when compared with that derived from the comparison sample of isolated QSOs,  as shown in   Figure \ref{Gnohost1}.
We find that the mean galaxy overdensity around QSO in pairs within 0.25 Mpc   
is $<$$G_{0.25}$$>$ = 1.45$\pm$0.07  ($<$$G_{0.25}$$>$ =  1.33$\pm$0.08 
if the companion host galaxy is excluded), 
while for the comparison sample  we obtain $<$$G_{0.25}$$>$ = 1.13$\pm$0.04.  
The increase of overdensity in the case of paired QSOs at $r$=250 kpc is  $\sim$ 25 \%. 
   
At ten kpc scales QSOs appear  more clustered  
\citep[e.g.][]{Hennawi2006,Myers2007,Myers2008,Kayo2012,Eftekha2017} 
 compared to the power-law extrapolation of clustering measurement at larger scales 
 \citep[Mpc, e.g.][]{Porciani2004}. This may be interpreted as an indication of dissipative interactions 
\citep{Djorgovski1991,Kochanek1999,Myers2007} in a comparable-scale richer environment.
 Intriguingly,  the immediate galaxy environment of the very close pairs in our sample (R$_{\bot}<$ 30 kpc) 
 appears depleted.
No  galaxy is detected in the SDSS images up to projected distances $\la$ 80 kpc.
If we exclude the companion QSOs (to limit their influence on the cumulative numbering 
of the surrounding poor environment),  the cumulative galaxy overdensity
profile reveals a underdensity  ($G$=0.5-1)
 with respect to the background extending up to $\sim$100-150 kpc, 
 followed by flat  trend where $G(r)$ $\sim$ 1. 
  If we focus on the entire environment surrounding both QSOs by measuring the overdensity
   from the midpoint of the pair, we find that it  appears on average still underdense compared 
   both to that of the remaining pairs and that of isolated QSOs up  to distances of $\sim$ 300 kpc.
However,  it is worth noting that this result represents only an indication, 
since it is based on a very small number of QSOs.

\section{Nearest neighbour galaxy analysis}\label{thenng}

We  examine the SDSS photometrically classified galaxies   in the immediate vicinity of  our sampled QSOs 
 searching for possible links between their  properties and the QSO activity.
In Figure  \ref{pdcomp}  we compare the distributions of the projected distance 
of the 1$^{st}$-nearest neighbour   galaxy of paired QSOs with that of isolated ones. 
When  the companion-QSO host galaxies are taken into account 
   the 1$^{st}$-nearest neighbour  galaxies are found globally closer to paired QSOs
(mean projected distance 61.0 kpc), with respect to isolated ones (mean 88.4 kpc), 
 whose distribution extend for nearly twice the distance.
About two thirds of 1$^{st}$-nearest neighbours  of the QSOs in pairs (24 out of 40) are located 
at distances $<$ 60  kpc, where strong interactions may be expected. 
Nearly half  of them (11 out of 40) are the companion-QSO hosts in the closest pairs.
The other ones, $\sim$ 33\% of the total sample,
are  inactive galaxies, mainly concentrated around 50 kpc from the QSO.
This same fraction (69 out of 200) of closest companion galaxies is detected around 
single QSOs at these distances. The luminosity distributions of the 1$^{st}$-nearest neighbours 
 within 60  kpc range from $-$20 to $-$26 mag for both the samples, with differences towards higher luminosities 
 in the case of paired QSOs, depending only on the host galaxy contribution. 

In conclusion, the exclusion of the hosts from the search of the 1$^{st}$-nearest
 neighbours makes  the distance and luminosity distributions around QSOs in pairs  
 indistinguishable from those around  isolated QSOs.

 \begin{figure}
\includegraphics[trim=0cm 0cm 0cm 0.0cm,clip,width=1\columnwidth]{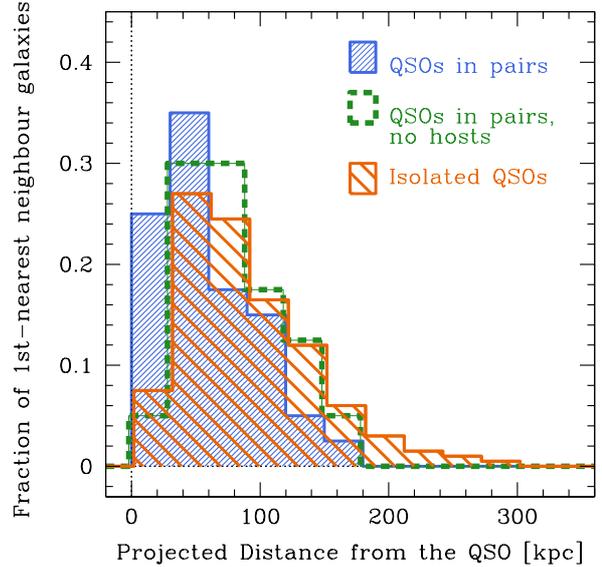}
   \caption{ \label{pdcomp} 
 	 Normalized distributions of the projected distances from the QSOs of 
	1$^{st}$-nearest neighbour galaxies brighter than the magnitude threshold 
 	$m_{i,50\%}$, observed in the SDSS i-band images. 
	Filled (blue) and dashed-contour (green) histograms
	refer, respectively,  to the case of inclusion or exclusion 
	of the host galaxy of the companion QSO  from the 1$^{st}$-nearest neighbour galaxy search.
	The distribution around isolated QSOs is reported as the line-filled (orange) histogram.
	}
  \end{figure}

   \section{Discussion}\label{thediscussion}

We selected 20 low redshift (z$_{ave}$ $\sim$ 0.5) likely physical QSO pairs,
with the aim of characterizing the galaxy  environment  around each QSO
and investigate if there is a  connection  between the environment properties 
and the QSO activity.
These properties are compared to those of  a consistent sample of isolated QSOs
at the same redshifts  and with similar global and host galaxy luminosities.
 Our results indicate that the galaxy environment of  low redshift QSOs that are in pairs  
 is, on average, richer than that around isolated QSOs (Figure \ref{Gnohost1}). 
Moreover, based on the present data, there is a suggestion that the scenario 
could be more complex, depending on the pair separation. 
The comparison of our current results with previous studies  on the QSO pair galaxy environment  
\citep{Fukugita2004,Boris2007,Farina2011,Green2011,Sandrinelli2014,Onoue2017}
does not help to improve the understanding of the picture because 
of the relevant differences among these works (e.g. differences of redshift, QSO luminosity, pair-member 
separation, difference of radial velocity, deepness of observation, etc.)

A   local and  overdense environment  of galaxies 
could have a role  on the generation of powerful QSO  nuclear activity.
The link between the richness of galaxy groups and the presence of an active nucleus 
could arise just by the higher probability of interaction in richer environments. 
However, it does not imply  that in rich clusters of galaxies one expects to find 
 many QSOs \citep[e.g.][]{Kauffmann2004,Coldwell2006}. 
The most likely environments for interactions are poor groups 
\citep[e.g.][] {Silverman2009},
in which galaxies have low relative velocities \citep[e.g.][]{Ostriker1980,Kauffmann2004,Popesso2006}
and  more cold gas content \citep[e.g.][]{McNamara2007}.
  This is  also consistent with  models on the cosmological role of QSOs 
 \citep[e.g.][]{Hopkins2008},  where major mergers between  gas-rich galaxies 
 are expected to preferentially occur in small-scale clustering excess.
Our findings,   although based on a small sample at low redshift,  appear in agreement
 with the role of major merger in small group of galaxies for sustaining  QSO activity \citep[e.g.][]{Hopkins2008}. 

 We note that in the cases of closest QSO pairs (projected separations $<$30 kpc) 
 the detection of a less dense environment is consistent with the fact that around these 
 QSOs there is no evidence of extended  X-rays emission \citep{Green2011}.
The interpretation of the poor environment  around very close QSO pairs
is not trivial, since the probability of interactions should be enhanced by clustering 
excess on small scales.
 We may argue that these cases represent a situation where the nuclear activity 
is triggered by the mutual/contemporaneous interaction of two massive galaxies, or possibly also with 
other group-companion galaxies (see the case of QSO pair QP18 in Appendix \ref{appA}).
If proved, the  tendency of these rare QSO pairs  to live in particularly modest environments 
may be supported by some kind of suppression effects, 
similar to those invoked in the models of e.g.  \cite{Bonoli2009} and \cite{Fanidakis2013},
where the QSO activities in haloes more massive then  $\sim$10$^{12}$-10$^{13}$ M$_{\odot}$ 
are inhibited due to the suppression of gas cooling by active galactic nucleus feedback.

As the general indication is that pairs inhabit small groups of galaxies extending up to few hundreds kpc
rather than clusters, at least in the redshift range of our investigation,  they cannot be used for 
locating large-scale structures, as suggested by \cite{Djorgovski1999} for the case of  high redshift QSOs,  
even if more than one  QSO may co-exists in massive structure \citep[e.g.][]{Onoue2017}.

Our results and interpretations could be probed using a  larger sample of QSO pairs,  
extending also at larger z, and an adequate comparison  of a suitable sample of isolate QSOs.
A better characterization of the environment   is now  feasible with large 
 telescopes equipped with Multi-Object Spectroscopy at  optical and near-IR wavelengths.
 Physical evidences of whether and which galaxies are associated with QSOs, 
their  velocity dispersion and  dynamical mass  compared with the expectation
from models may be achieved.
The concurrent  search for environmental galaxy signatures of recent star formation  through 
the  H$\alpha$ emission line will allow us  to fully explore/study the environment-activity relations.

\section*{Acknowledgements}

We thank the anonymous referee for  useful comments and suggestions which enabled
us to improve the paper.

Quasars in our QSO pair sample were published as the Half Million Quasars 
 catalogue \citep[versions 3.9-4.3,][]{Flesch2015}. 

Funding for the Sloan Digital Sky Survey IV has been provided by the Alfred P. Sloan 
Foundation, the U.S. Department of Energy Office of Science, and the Participating Institutions. 
SDSS-IV acknowledges
support and resources from the Center for High-Performance Computing at
the University of Utah. The SDSS web site is www.sdss.org.

SDSS-IV is managed by the Astrophysical Research Consortium for the 
Participating Institutions of the SDSS Collaboration including the 
Brazilian Participation Group, the Carnegie Institution for Science, 
Carnegie Mellon University, the Chilean Participation Group, the French Participation 
Group, Harvard-Smithsonian Center for Astrophysics, 
Instituto de Astrof\'isica de Canarias, The Johns Hopkins University, 
Kavli Institute for the Physics and Mathematics of the Universe (IPMU) / 
University of Tokyo, Lawrence Berkeley National Laboratory, 
Leibniz Institut f\"ur Astrophysik Potsdam (AIP),  
Max-Planck-Institut f\"ur Astronomie (MPIA Heidelberg), 
Max-Planck-Institut f\"ur Astrophysik (MPA Garching), 
Max-Planck-Institut f\"ur Extraterrestrische Physik (MPE), 
National Astronomical Observatories of China, New Mexico State University, 
New York University, University of Notre Dame, 
Observat\'ario Nacional / MCTI, The Ohio State University, 
Pennsylvania State University, Shanghai Astronomical Observatory, 
United Kingdom Participation Group,
Universidad Nacional Aut\'onoma de M\'exico, University of Arizona, 
University of Colorado Boulder, University of Oxford, University of Portsmouth, 
University of Utah, University of Virginia, University of Washington, University of Wisconsin, 
Vanderbilt University, and Yale University.








\appendix

\section{GTC Optical Spectra}\label{appA}

 In Figures  \ref{gtcex} and \ref{gtc1} we show the optical spectra of QSO pairs obtained at GTC. 
Some notes on individual cases are reported in the following.

 \begin{figure*}
  \setcounter{figure}{0}
  \includegraphics[width=1\columnwidth]{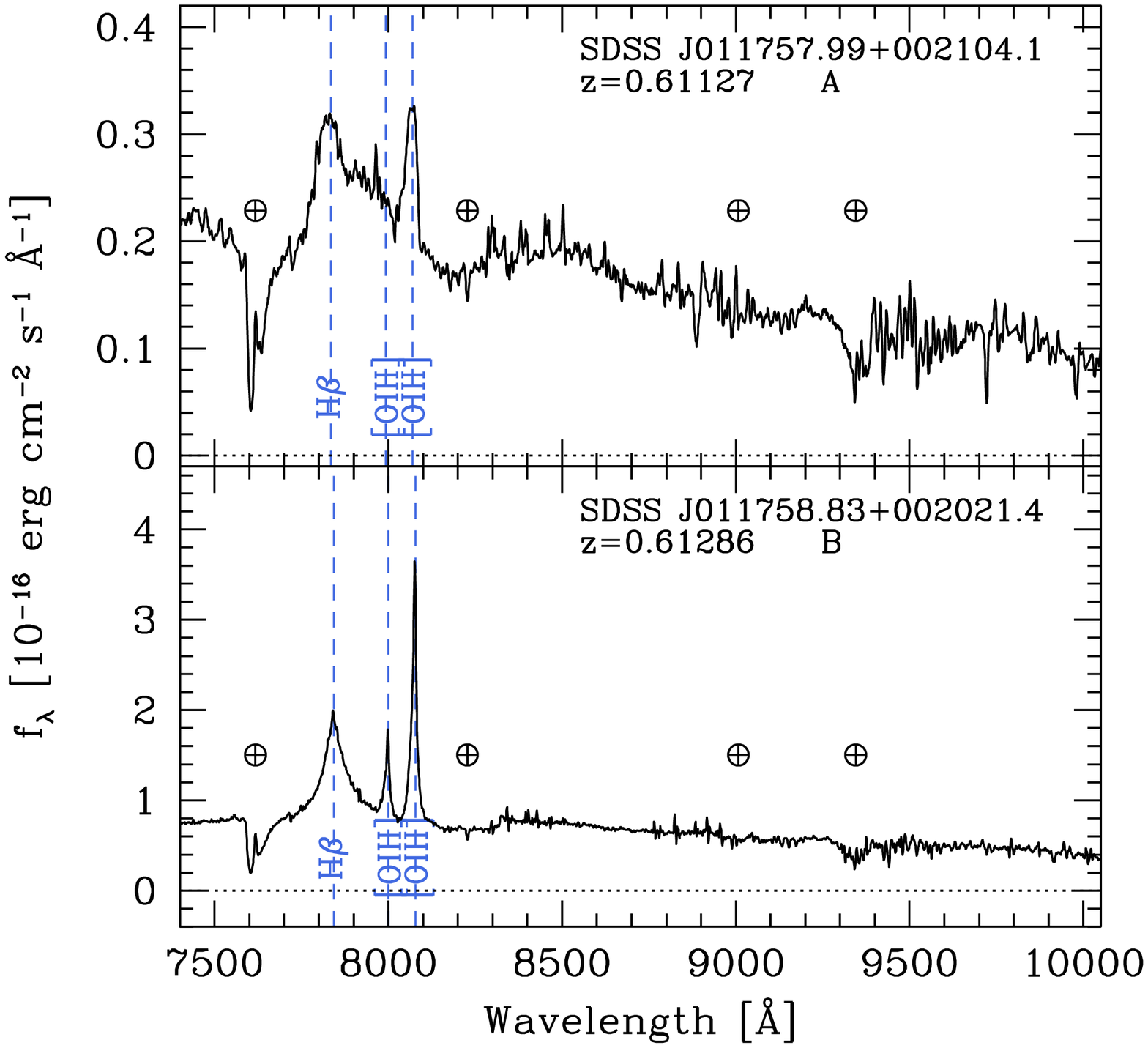}
 \vspace{-0.7cm}
 \includegraphics[width=1\columnwidth]{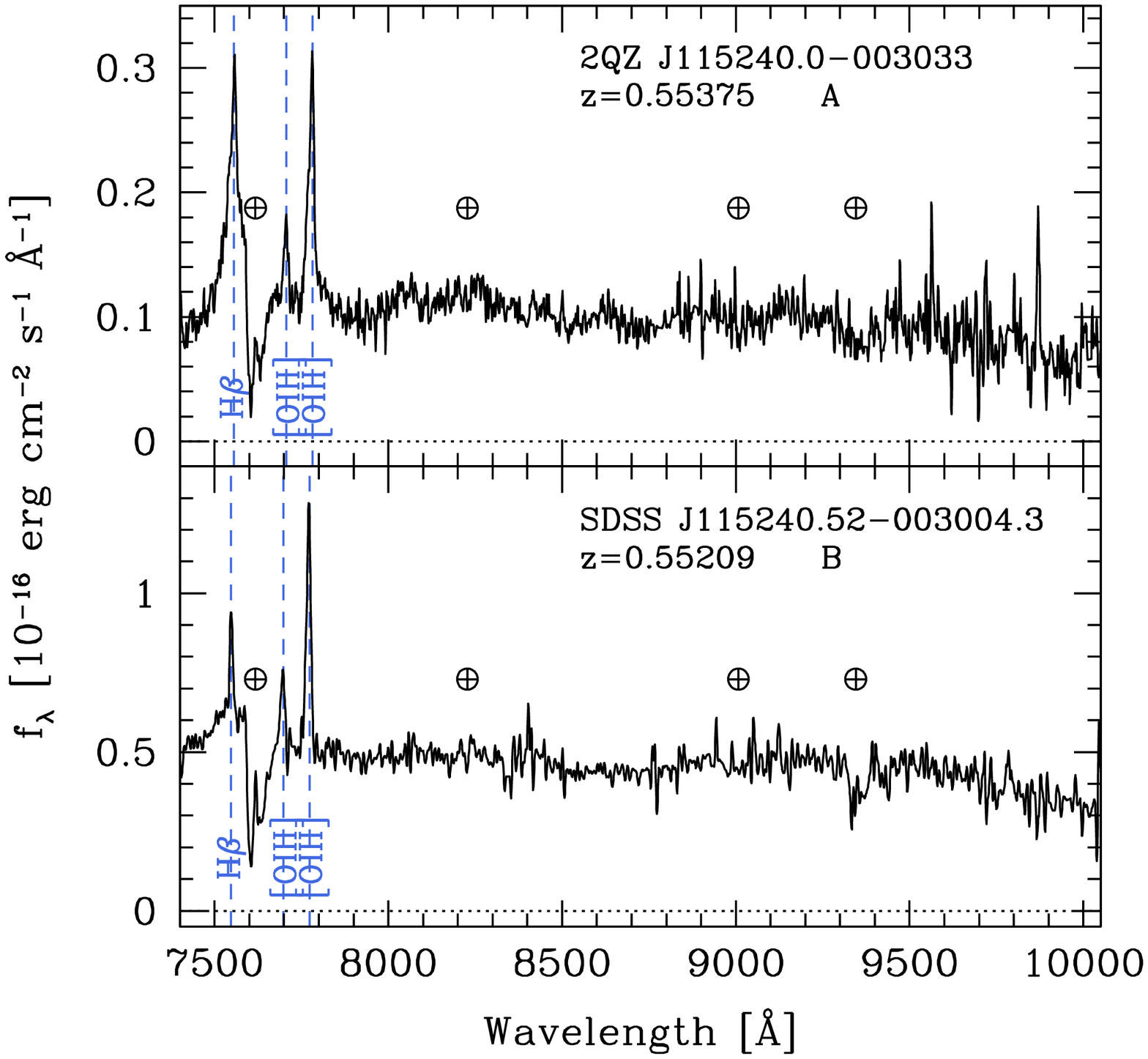}
\includegraphics[width=1\columnwidth]{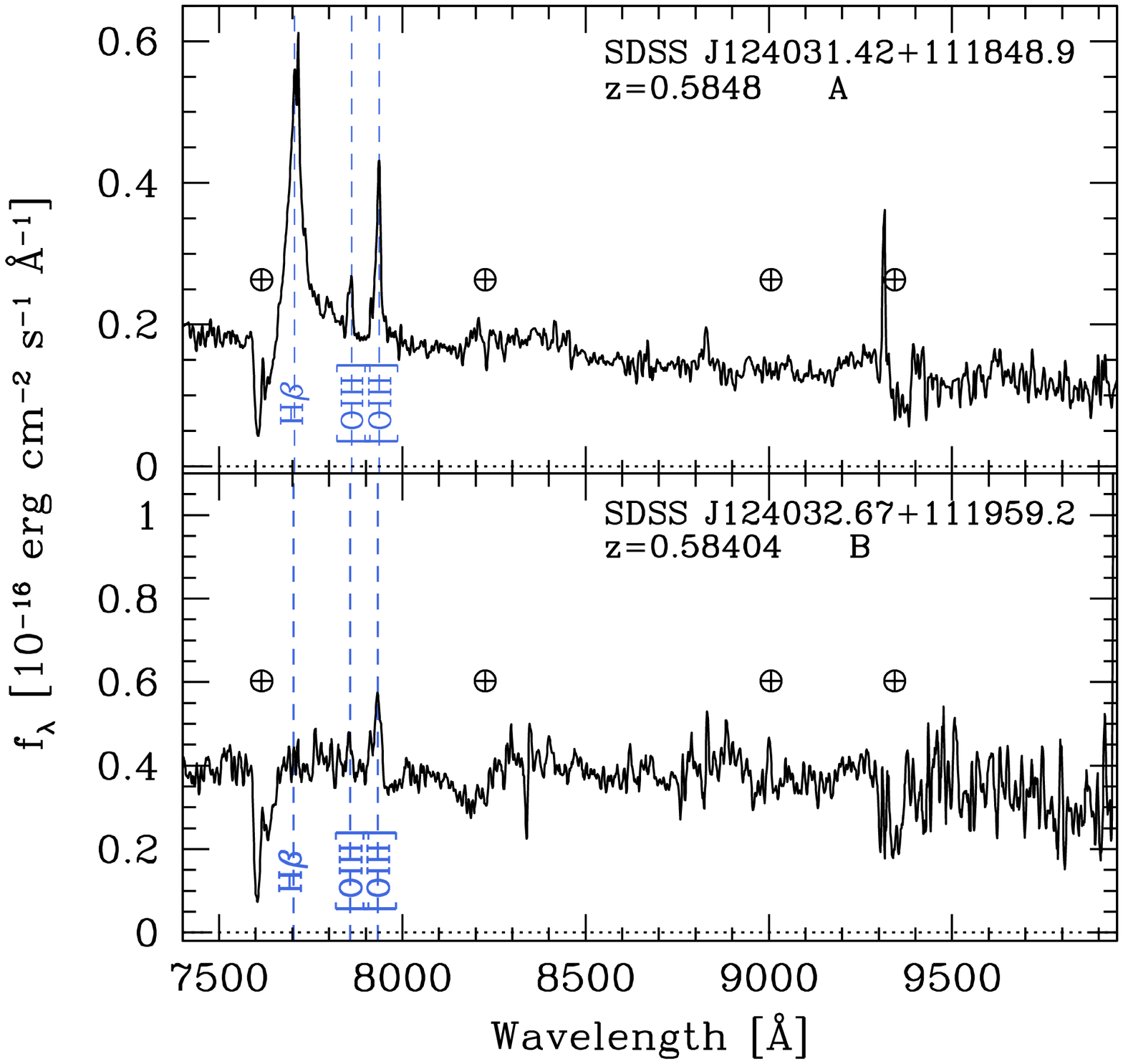}
 \hspace{-0.2cm}
 \includegraphics[width=1\columnwidth]{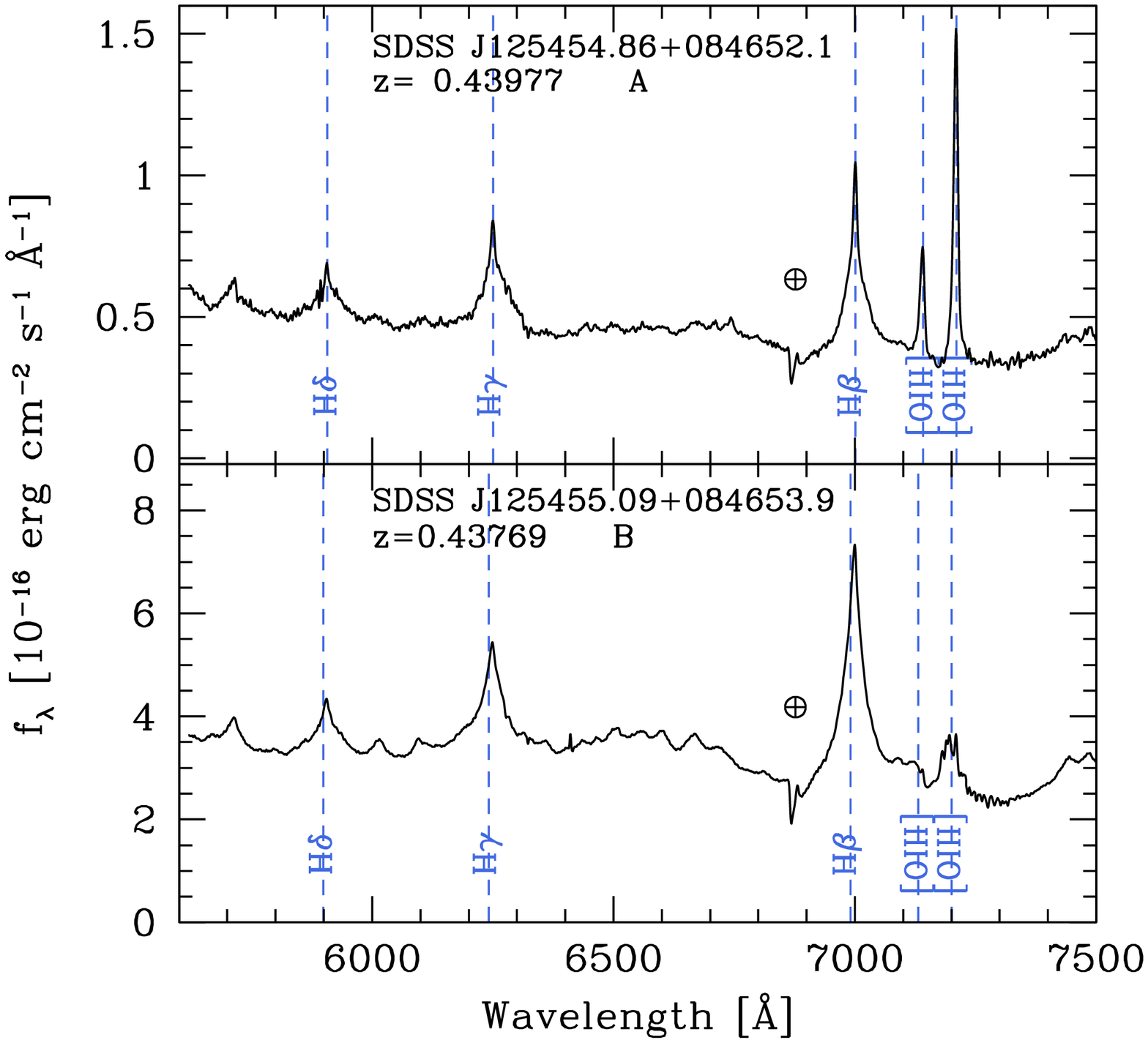}
  \caption{\label{gtc1}
 Spectra of the QSO physical  pairs observed at GTC.
The most prominent  emission features  are  marked. 
The main telluric bands are indicated by $\oplus.$
   }
\end{figure*}

 \begin{figure*}
  \setcounter{figure}{0}
 \includegraphics[width=1\columnwidth]{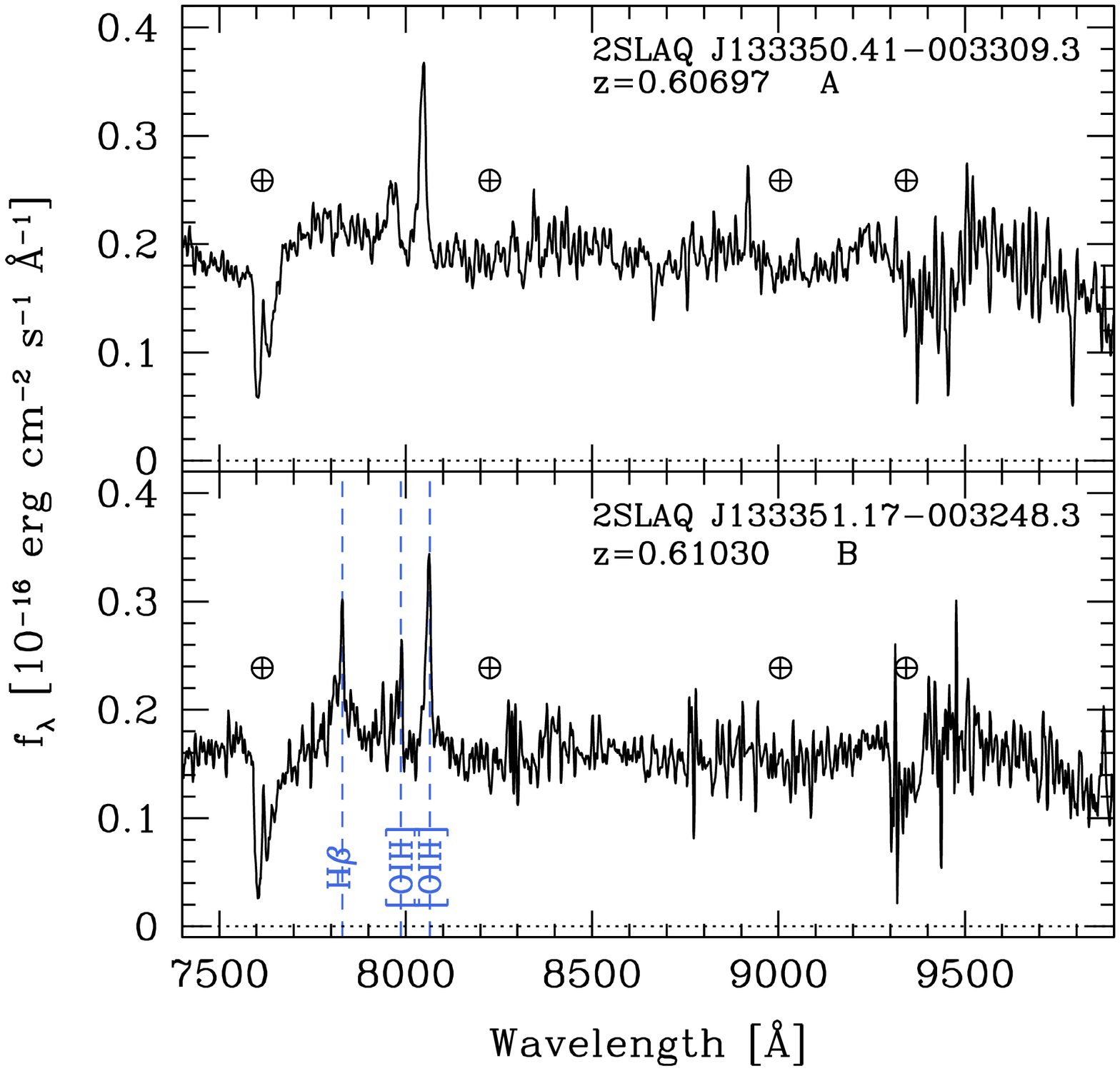}
 \vspace{-0.7cm}
 \includegraphics[width=1\columnwidth]{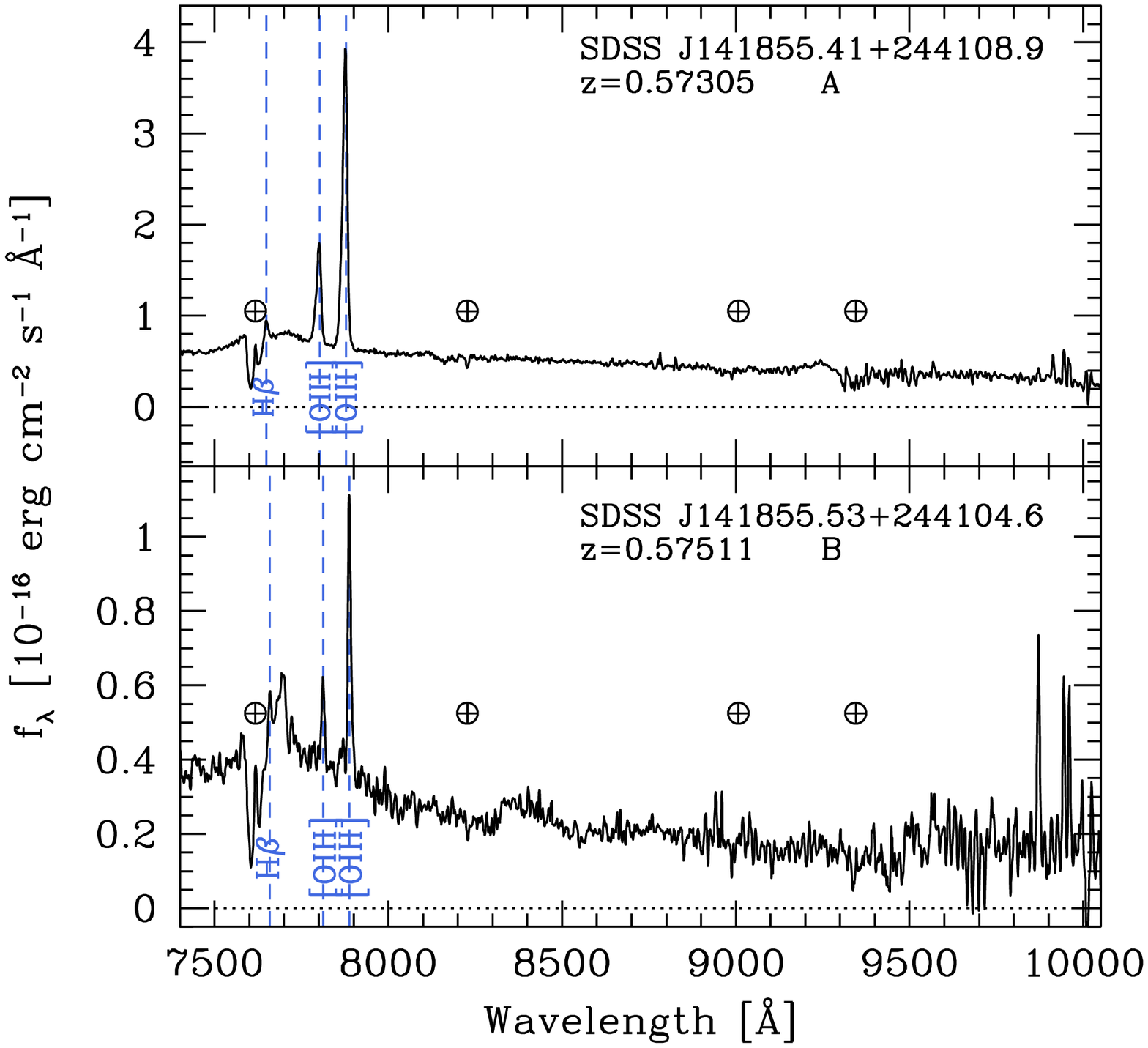} 
   \caption{\label{ctc2}
 --- Continued}
\end{figure*}

 \begin{figure}
  \setcounter{figure}{0}
 \vspace{-0.7cm}
\hspace{0.3cm}
  \includegraphics[width=1\columnwidth]{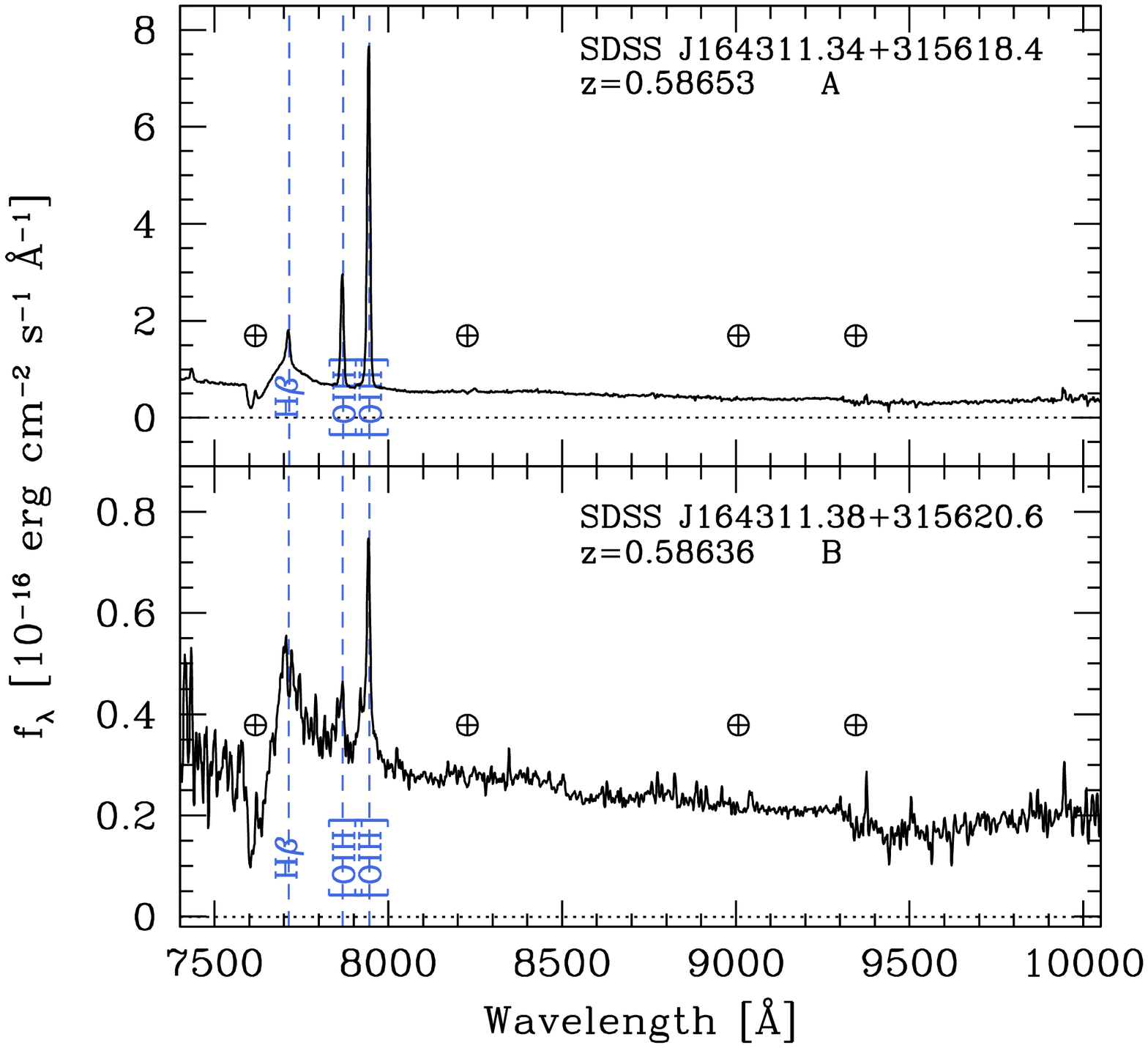} 
 \vspace{-0.7cm}
    \caption{\label{ctc2}
 --- Continued}
\end{figure}

\textit{J115822.77$+$123518.5,  J115822.98+123520.3 (QP13)} - 
Found by using the kernel density estimation  \citep[KDE,][]{Richards2004}, this QSO pair
 was  spectroscopically investigated   by \cite{Myers2008} as a close binary  QSO candidate. 
Due to the spectral resolution, component separation and similar colors,  \cite{Myers2008} 
didn't exclude  it was  a most likely lens candidate.  It was, however, recovered as a double QSO 
by  \cite{Foreman2009} and by  \cite{Green2011}. 
 Dissimilarities in the  peak-to-continuum ratios in our GTC spectra, reported in Figure \ref{gtcex}, 
 validate the physical nature of the QSO pair.
 
\textit{J124031.42$+$111848.9, J124032.67+111959.2 (QP14)} - 
Both the  spectra are available in the SDSS archives \citep{Schneider2010,Paris2014}, but
for the B component the  [OIII] line positions are not  measurable. 
On the GTC spectrum (Figure \ref{gtc1}), a  line  emerges in the position ascribable to
the [O III]${\lambda5007}$ when compared with  the broad MgII line and  other minor 
 emissions in the SDSS spectrum.  H$\beta$  line appears suppressed in both GTC 
 and  SDSS observations. 

\textit{J125454.86$+$084652.1,  J125455.09+084653.9 (QP 16)} -
It  was spectroscopically identified as close  partially resolved binary QSO by \cite{Green2010}, who 
 found it hosted in an ongoing  galaxy merger with clear tidal tail features. 
 Its environment was investigated by \cite{Green2011} in a multi-wavelength study of 
 binary QSOs, with other QSO pairs of our sample (see  Section \ref{theintro}).
 It was included in our GTC program with the aim to better measure the $\Delta V_\parallel$  
 through a reliable redshift measure of the B companion.
   As shown in Figure \ref{gtc1}, narrow lines are significantly  higher in the A component,
while [OIII] lines appear broken off  or absent at all,  similarly to the spectrum  observed 
by \cite{Green2010} on 2009 May 22.
 We note that, in comparison, in  the spectrum  retrieved from the SDSS archives and  
 observed on 2007 June15, the  [O III]${\lambda5007}$ line appears stronger  
 and the spectral continuum enhanced.

\textit{J141855.41$+$244108.9, J14189+2441B (QP 18)} -
It was identified as a close QSO pair ($\Delta \theta$=4.5\arcsec) by  \cite{Myers2008},
who tended to excluded a single-lensed source due to the low probability associated with 
pairs  with $\Delta \theta  >$3\arcsec, following \cite{Hennawi2006}.
However,  
they did not rule out the lens interpretation for this pair on the basis of their spectroscopy.  
 The pair was included in the QSO binary samples of \cite{Foreman2009},  \cite{Green2011} and 
 \cite{Eftekha2017}.
Only the spectrum of  the QSO A was recovered in literature (SDSS archives).
Due to spectral dissimilarities also in this case we confirm the physical nature  the QSO 
association. In the GTC 2D spectrum we note the presence of a companion galaxy not 
detected by the SDSS photometry, which is very close to the weaker  
component of this pairs (QSO B, see Figure {\ref{bidimspec}). 
 It is visible at the  [O III]${\lambda5007}$ and  [O III]${\lambda4959}$ emission line  positions
at approximately  the same redshift of the near QSO, and probably in interaction.
Its emissions were carefully separated from the QSO spectrum during the extraction phase.

\textit{J164311.34$+$315618.4, J164311.38+315620.6 (QP 20)} -
It is one of the closest-known QSO pairs, with projected separation 2.3\arcsec corresponding to 15 kpc.  
Discovered as a binary radio-loud/radio-quiet binary QSO by \cite{Brotherton1999}, 
it contains the only one object detected as radio-loud in our QSO pair sample (FIRST J164311.3+3156184).
By using multi-band imaging, \cite{Kunert2011}  found that the host galaxy of the radio-loud component
 is highly disturbed. They also observed an  intermittent  activity  of radio structure, possibly 
 due to the  rapid change of the jet direction and/or  to the interaction with the companion. 

\begin{figure}
 \centering
\vspace{2.3cm}
  \includegraphics[angle=270,width=0.8\columnwidth]{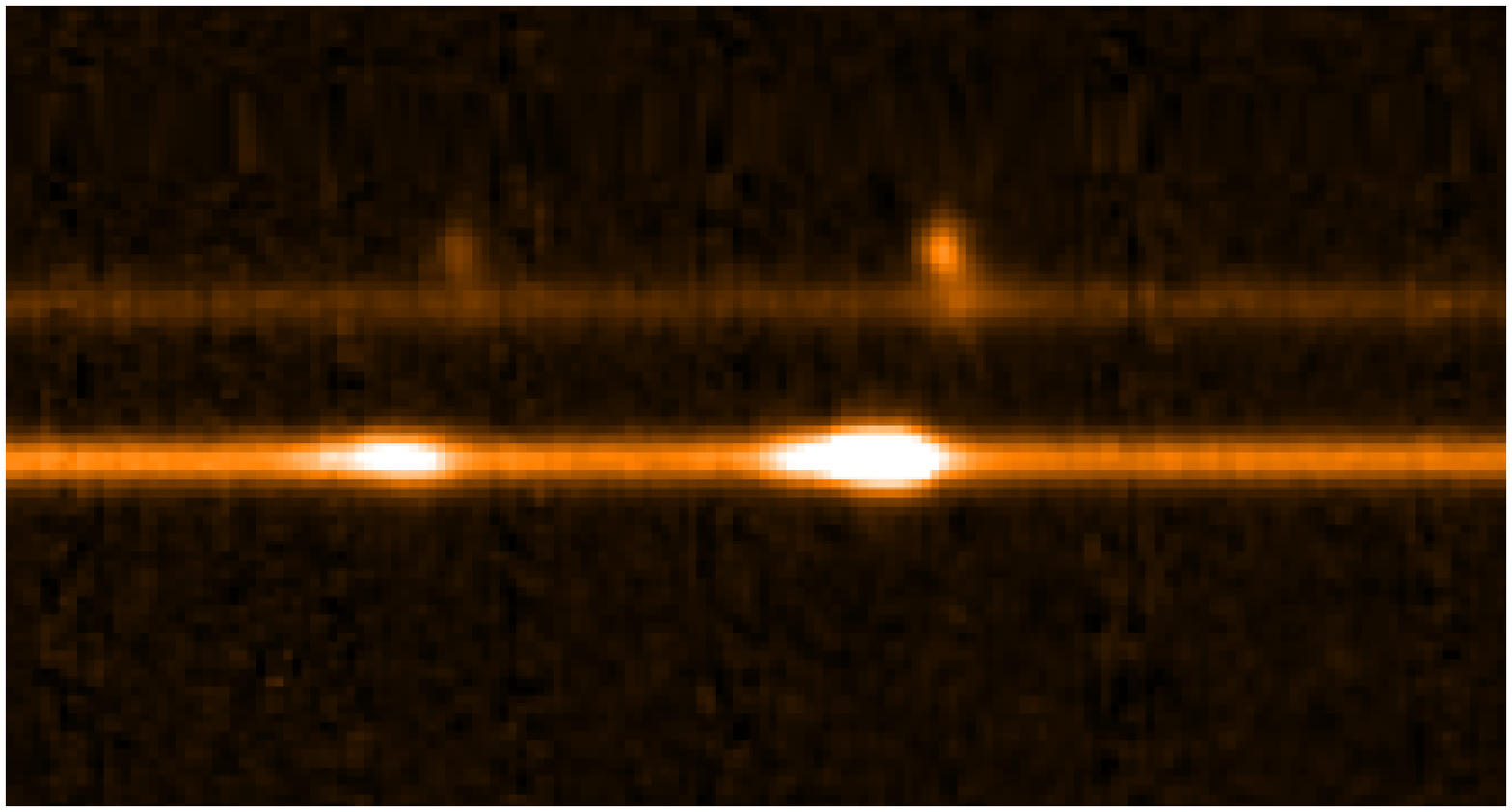}
\vspace{0.7cm}
    \caption{ \label{bidimspec} 
 GTC 2D spectra of the QSO pair QP18. The spectrum of  QSO A   is the brighter one.
At $\sim$ 1.4 arcsec ($\sim$ 9 kpc projected distance) from the spectrum of  QSO B,    
a  galaxy is apparent at the  [O III]${\lambda5007}$ and  [O III]${\lambda4959}$ 
emission line  positions at approximately the same redshift  of the near QSO. 
}
\centering
  \end{figure}

\section{Host galaxies}
{\label{appC}}
 
Galaxies hosting QSOs were photometrically  studied with the aim of 
outlining their properties and drawing  a comparison with  those  hosting isolated QSOs. 
The QSO images are  drawn in  i-band  from SDSS-DR12 imaging archives (see Subsection \ref{theim}).
We used the Astronomical Image Decomposition and Analysis \citep[\textsc{AIDA},][]{Uslenghi2008}
software  to model and  to decompose the host galaxy luminosity from the nuclear source. 
The nucleus  is described by a local Point Spread Function (PSF) generated through 
close stars in the image.  
Host halo may be resolved by a second fit procedure if the luminosity profile results 
best fitted by the galaxy model represented by a \cite{Sersic1963} law convolved 
with the proper PSF. 
 The two fit outputs are compared by the  $\chi^2_{PSF} / \chi^2_{PSF+host}$  
 ratio  and  visually inspected. Example of fits are given in Figure \ref{AIDA}.
  The  employed  technique  was applied in  previous 
 QSO host galaxy studies   \citep[e.g.][]{Falomo2008,Decarli2012,Sandrinelli2014}  
 and widely discussed in \cite{Falomo2014}.

\begin{table*}
\begin{center}
\caption{Properties of nuclei and host galaxies of the QSO pairs.}
\label{AIDAtable}
\begin{tabular}{@{}llcccc|llcccc}
\hline\hline
QSO	&Class	&i$_{nuc}$ & i$_{host}$ &M(r)$_{nuc}$ &M(r)$_{host}$ &QSO	&Class &i$_{nuc}$ & i$_{host}$ &M(r)$_{nuc}$ &M(r)$_{host}$\\
	&	&[mag]	   & [mag]      &[mag]        &[mag]	     &		&      &[mag]     & [mag]      &[mag]        &[mag]	\\
(a)	& (b)	&(c)	   &(d)	        &(e)	      &	(f)	     & (a)      & (b)  &(c)	  &(d)	       &(e)	     & (f)	\\
\hline          
&&&&&&&&&\\  
1A	& M     & 19.90    & 21.61      & $-$22.71    & $-$21.17     & 11A  	& R    & 20.50    & 19.98      & $-$22.85    & $-$23.75 \\ 		
1B	& R     & $>$22    & 20.35      & $>$ $-$21   & $-$22.42     & 11B 	& R    & 20.28    & 21.23      & $-$23.09    & $-$22.51 \\ 
2A	& R     & 20.21    & 20.55      & $-$22.98    & $-$22.98     & 12A  	& U    & 20.00    &   ---      & $-$22.93    &      --- \\ 
2B	& U     & 18.23    &   ---      & $-$24.96    &    ---       & 12B  	& R    & 19.32    & 18.84      & $-$23.60    & $-$24.33 \\ 
3A	& R     & 21.68    & 19.51      & $-$20.61    & $-$22.86     & 13A  	& R    & 19.63    & 20.33      & $-$23.50    & $-$23.10 \\ 
3B	& R     & 20.16    & 19.03      & $-$22.13    & $-$23.33     & 13B  	& M    & 20.42    & 20.04      & $-$22.72    & $-$23.39 \\ 
4A	& R     & 20.72    & 20.63      & $-$21.92    & $-$22.17     & 14A  	& U    & 19.96    &   ---      & $-$23.22    &      --- \\ 
4B	& M     & 18.55    & 20.09      & $-$24.09    & $-$22.71     & 14B  	& R    & 20.75    & 19.64      & $-$22.42    & $-$23.81 \\ 
5A	& R     & 19.19    & 19.55      & $-$24.26    & $-$24.30     & 15A  	& M    & 19.35    & 19.34      & $-$22.98    & $-$23.09 \\ 
5B	& M     & 19.45    & 19.90      & $-$23.28    & $-$23.00     & 15B  	& U    & 18.30    &    ---     & $-$24.02    &      --- \\ 				
6A	& U     & 19.94    &   ---      & $-$22.74    &    ---       & 16A  	& R    & 19.76    & 19.21      & $-$22.69    & $-$23.35 \\ 
6B	& M     & 20.85    & 20.44      & $-$22.52    & $-$23.32     & 16B  	& U    & 16.97    &   ---      & $-$25.48    &      --- \\ 
7A	& R     & 18.90    & 19.76      & $-$24.00    & $-$23.35     & 17A  	& M    & 20.22    & 20.52      & $-$23.05    & $-$23.07 \\ 
7B	& M     & 18.77    & 20.43      & $-$24.14    & $-$22.69     & 17B  	& R    & 21.25    & 20.19      & $-$22.03    & $-$23.43 \\ 
8A	& M     & 18.83    & 20.70      & $-$24.09    & $-$22.44     & 18A  	& U    & 18.92    &   ---      & $-$24.11    &      --- \\ 				
8B	& U     & 18.61    &   ---      & $-$24.30    &    ---	     & 18B  	& M    & 20.10    & 20.80      & $-$22.94    & $-$22.50 \\ 	  	
9A	& R     & 18.68    & 19.48      & $-$24.20    & $-$23.63     & 19A  	& R    & 18.29    & 20.04      & $-$25.13    & $-$23.77 \\ 
9B	& U     & 19.17    &   ---      & $-$23.72    &    ---       & 19B  	& U    & 20.64    &    ---     & $-$22.79    &      --- \\ 
10A	& U     & 19.30    &   ---      & $-$23.76    &    ---       & 20A  	& R    & 19.38    & 19.06      & $-$23.71    & $-$24.32 \\ 
10B	& U     & 19.95    &   ---      & $-$23.10    &    ---       & 20B      & R    & 19.59    & 19.38      & $-$23.49    & $-$23.99 \\	
&&&&&&&&&\\  
\hline
 \end{tabular}
\end{center} 
\begin{tablenotes}
\item Notes.  
(a) Quasar identifier: QSO pair number + QSO component.
(b)  Resolved (R),  marginally resolved (M), unresolved (U) host galaxy.
(c) and (d)  Apparent i-magnitude   of the nucleus and host galaxy.
 (e) and (f)  Absolute r-band magnitude (k-corrected and dereddered) 
 of   nucleus and host galaxy.
\end{tablenotes}
 \end{table*}

\begin{figure*}
  \includegraphics[trim=0.2cm 0cm 0.5cm 0cm,clip,width=0.69\columnwidth]{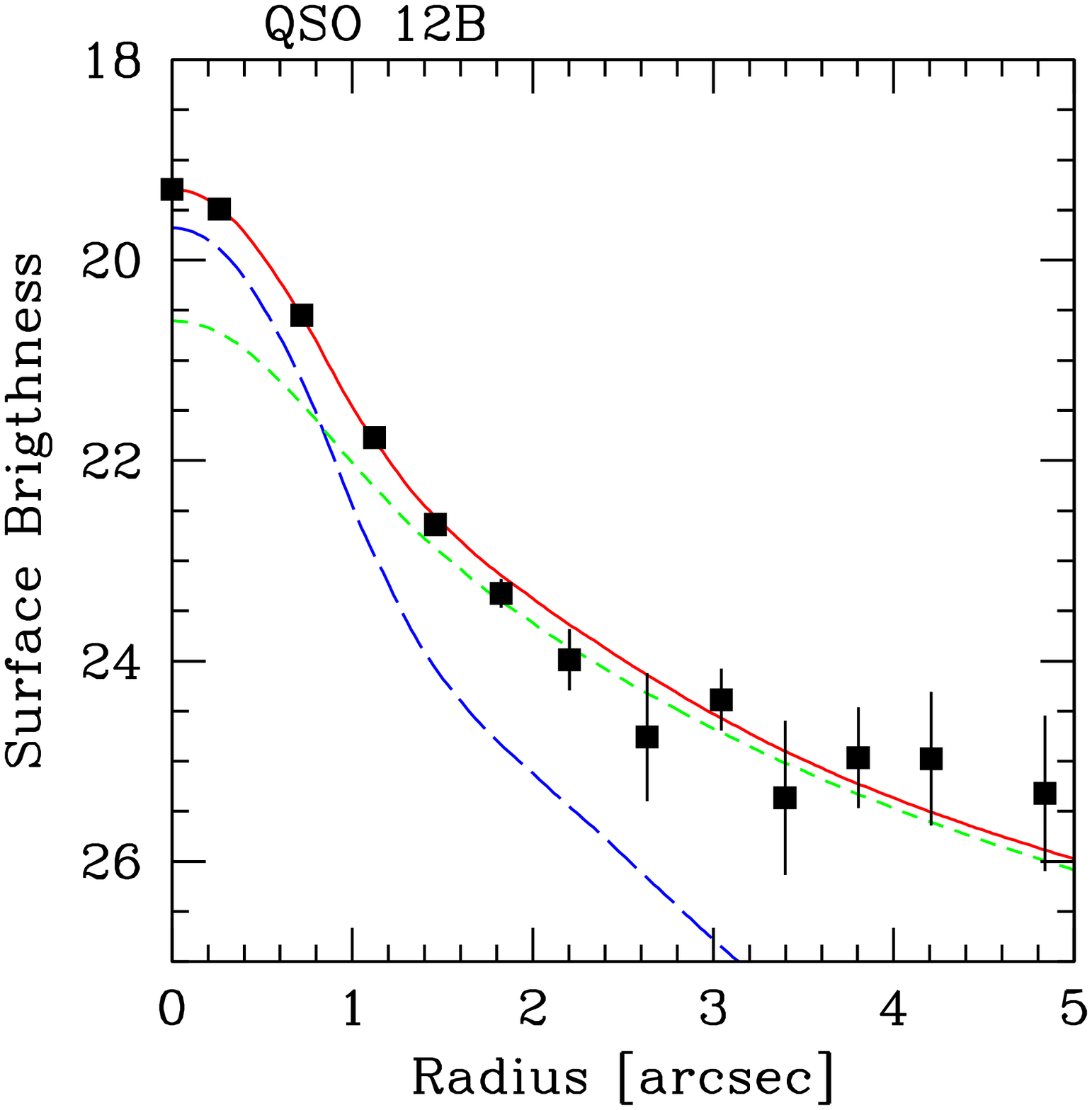}
  \includegraphics[trim=0.2cm 0cm 0.5cm 0cm,clip,width=0.69\columnwidth]{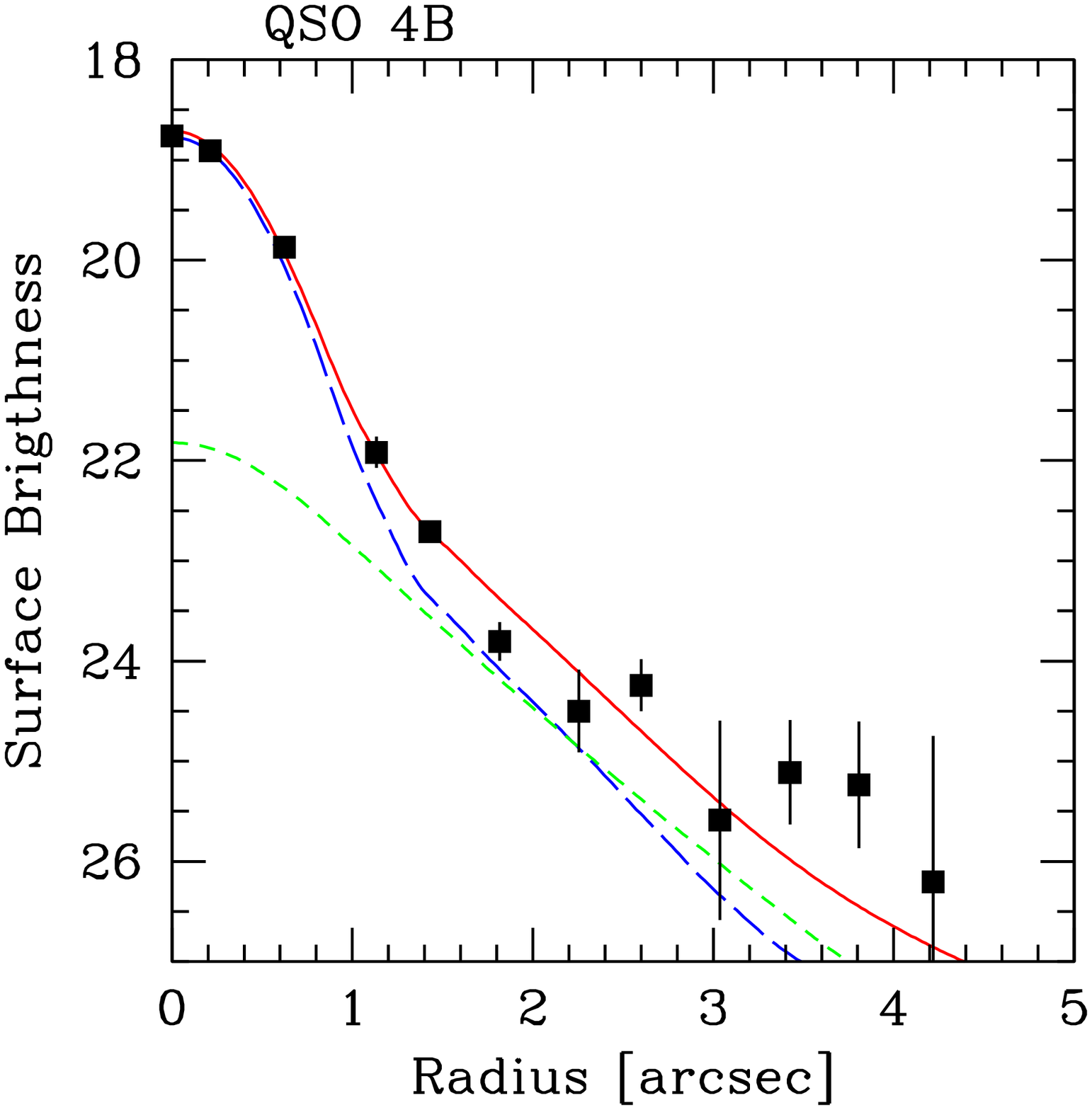}
  \includegraphics[trim=0.2cm 0cm 0.5cm 0cm,clip,width=0.69\columnwidth]{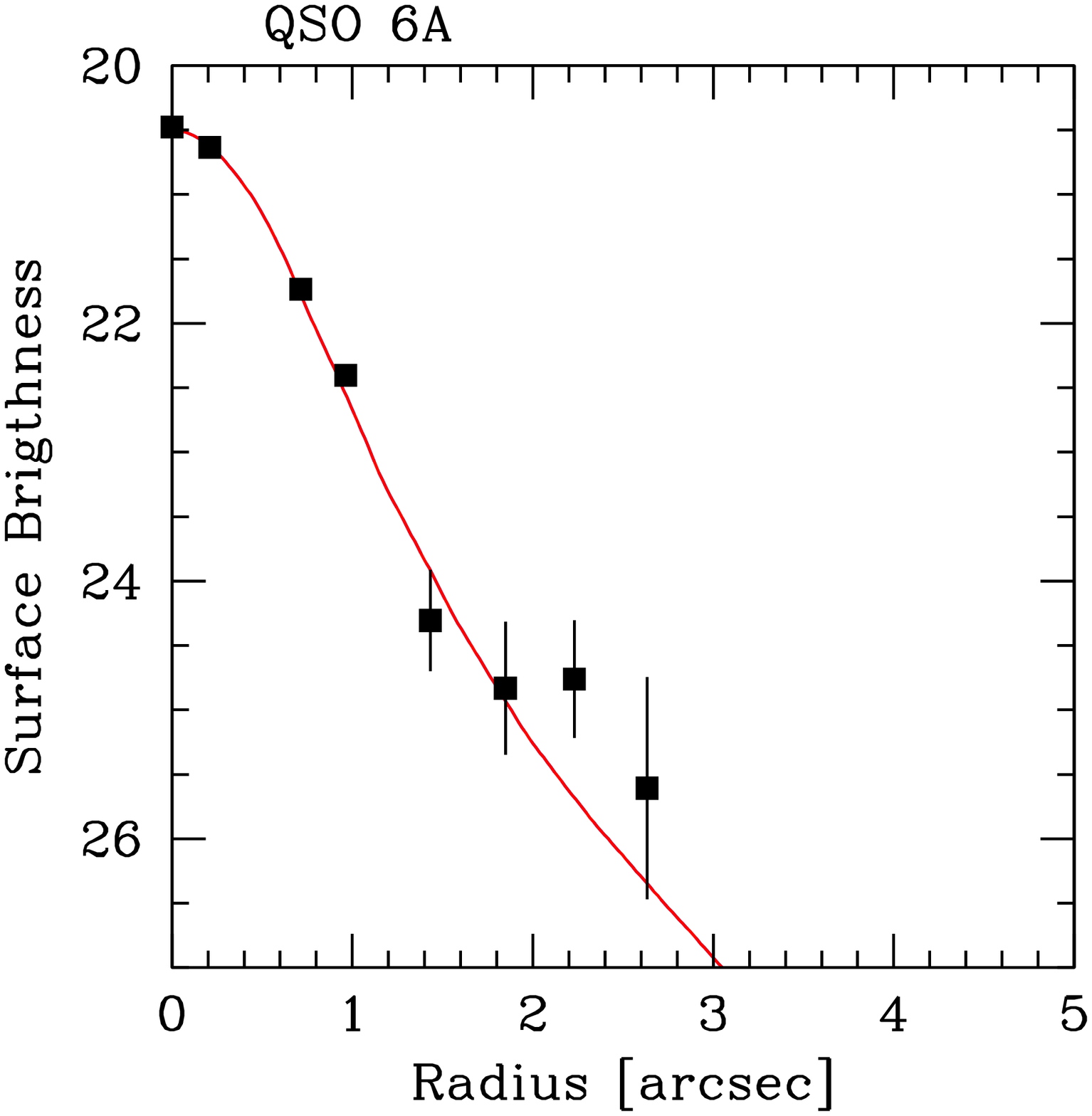}
    \caption{ \label{AIDA} 
    Examples of  the QSO luminosity  decomposition applied to QSOs with  resolved (QSO 12B),
    marginally resolved (QSO 4B) and unresolved host galaxies (QSO 6A).  
The average radial brightness profile of the QSO (square dots) is fitted by the 
 scaled PSF (blue dashed line) and the  
 host galaxy  Sersic law model  convolved with the PSF (green short dashed line).
The best fit is represented by the solid line
 }
  \end{figure*}

The classification results in 18  (45\%) resolved host galaxies (R), 
8 (20\%) marginally resolved (M), and 17 (33\%)  unresolved (U); 
for 9 pairs we are able to characterize the host galaxy properties of both QSOs.
Nucleus and host galaxy i-magnitudes are reported in Table \ref{AIDAtable}, 
together with the rest-frame absolute  SDSS r-magnitudes, dereddened and k-corrected. 
Corrections for galactic extinction were taken from SDSS data base.
K-corrections derive from templates of  \cite{Mannucci2001} and  \cite{Francis2001} 
 for host galaxies and nuclei, respectively. 
We find that no obvious differences  are apparent in the absolute magnitude  
distributions of the host galaxies of the two samples.  
 They range between M(r)$_{host}=$$-$21 mag and M(r)$_{host}=$$-$25.5 mag,  
 with the bulk of galaxies between M* and M*$-$2.
 Mean values  and medians  are similar, respectively 
   ($-$23.2$\pm$0.01) mag 
   and ($-$23.1$\pm$0.01) mag for paired QSOs and 
($-$23.4$\pm$0.01) mag 
and ($-$23.4$\pm$0.01) mag for those that are isolated.
  The indication is that the two families of QSO  are indistinguishable with respect  to the host
 luminosity.


\bsp	
\label{lastpage}
\end{document}